\documentstyle[preprint,eqsecnum,aps,epsf,tighten]{revtex}

\begin{document}

\newcommand{\dfrac}[2]{\frac{\displaystyle #1}{\displaystyle #2}}
\newcommand{\dss}{\delta s^2}
\newcommand{\dhL}{\delta h_L}
\newcommand{\dhdR}{\delta h_{d_R}^{\not R}}
\newcommand{\dhsR}{\delta h_{s_R}^{\not R}}
\newcommand{\dhbR}{\delta h_{b_R}^{\not R}}
\newcommand{\dhbLH}{\delta h_{b_L}^{\rm Higgs}}
%%%%%%%%%%%%%%%%%%%%%%%%%%%%%%%%%%%%%%%%%%%%%%%%%%%%%%%%%%%%%%%%%%%%%%%%%%%%%
\draft
\preprint{VPI--IPPAP--99--10}

\title{Constraints on R--parity violating couplings from
CERN LEP and SLAC SLD hadronic observables}
\author{
Oleg~Lebedev\thanks{electronic address: lebedev@quasar.phys.vt.edu},
Will~Loinaz\thanks{electronic address: loinaz@alumni.princeton.edu}, and
Tatsu~Takeuchi\thanks{electronic address: takeuchi@vt.edu}
}
\address{Institute for Particle Physics and Astrophysics,
Physics Department, Virginia Tech, Blacksburg, VA 24061}

\date{Revised January 15, 2000}
\maketitle

\begin{abstract}
We analyze the one loop corrections to hadronic $Z$ decays
in an R--parity violating extension to the Minimal Supersymmetric
Standard Model (MSSM).  Performing a global fit to all the
hadronic observables at the $Z$--peak, we obtain stringent
constraints on the R--violating coupling constants 
$\lambda^{\prime}$ and $\lambda^{\prime \prime}$.  
The presence of these couplings worsens the agreement with the data 
relative to the Standard Model.
The strongest constraints come from the $b$ asymmetry
parameters $A_b$ and $A_{\rm FB}(b)$.  From a {\it classical}
statistical analysis
we find that the couplings
$\lambda^{\prime}_{i31}$,
$\lambda^{\prime}_{i32}$, and 
$\lambda^{\prime\prime}_{321}$
are ruled out at the 1$\sigma$ level, and that $\lambda^{\prime}_{i33}$
and $\lambda^{\prime\prime}_{33i}$ are ruled out at the 2$\sigma$ level.
We also obtain Bayesian confidence limits for the R--violating couplings.
\end{abstract}

\pacs{12.60.Jv, 12.15.Lk, 13.38.Dg}

%%%%%%%%%%%%%%%%%%%%%%%%%%%%%%%%%%%%%%%%%%%%%%%%%%%%%%%%%%%%%%%%%%%%%%%%%%%%%%
\narrowtext

\section{Introduction}

R--parity conservation is often assumed in supersymmetric model 
building in order to prevent a host of phenomenological complications 
such as fast proton decay.
This also serves to make the lightest supersymmetric particle (LSP) stable   
and thus provide a dark matter candidate.
However, R--parity conservation is not a {\it necessary} condition for 
avoiding many of these problems. 
For example, the imposition of other discrete symmetries,
such as conservation of either baryon number or lepton number, may be 
adequate to provide phenomenologically acceptable models.  
(For recent reviews, see Ref.~\cite{Dreiner:1997uz}.)  
Furthermore, the evidence for neutrino mass recently observed at 
Super--Kamiokande \cite{SuperK:98} lends improved motivation to 
consider R--parity violating extensions to the minimal supersymmetric
standard model (MSSM).
Therefore, one is led to question just how much R--parity violation 
can be introduced without conflict with current experimental data.
In this paper, we study the radiative corrections from 
R--parity violating extensions 
of the MSSM to the electroweak observables in hadronic $Z$ decays, namely
the ratios of hadronic partial widths and the parity violating asymmetries.
Experimental data from LEP and SLD place stringent limits on the 
size of these corrections, thereby constraining the possible 
strengths of the R--violating interactions.

We focus on the effects of the R--parity violating superpotential 
and neglect possible effects from the corresponding 
soft--breaking terms \cite{PILAFTSIS:96}. 
This simplification allows us to rotate away the bilinear 
terms \cite{Hall:1984id}. 
In this case, the R--parity violating superpotential has the following form:
\begin{equation}
W_{\not{R}}
= \frac{1}{2}\lambda_{ijk}   \hat{L}_i \hat{L}_j \hat{E}_k 
+            \lambda_{ijk}'  \hat{L}_i \hat{Q}_j \hat{D}_k 
+ \frac{1}{2}\lambda_{ijk}'' \hat{U}_i \hat{D}_j \hat{D}_k\;,
\label{eq:superpotential}
\end{equation}
where $\hat{L}_i$, $\hat{E}_i$, $\hat{Q}_i$, $\hat{U}_i$, and $\hat{D}_i$ 
are the MSSM superfields defined in the usual fashion \cite{Haber:1985rc}, 
and the subscripts $i,j,k=1,2,3$ are the generation indices.  
These interactions can give potentially sizeable radiative corrections to 
the hadronic observables depending on the size of the coupling constants
$\lambda$, $\lambda^{\prime}$, and $\lambda^{\prime\prime}$.
The $\lambda$ couplings are already tightly 
constrained to be $O(10^{-2})$ or less, and their effect 
on the $Z$--peak observables is negligible \cite{Allanach:1999ic}.\footnote{%
They do not affect the quark couplings in any case.} 
The constraints on the $\lambda^{\prime}$ and $\lambda^{\prime\prime}$ 
couplings are much less stringent.  
However, they cannot be present simultaneously in the Lagrangian 
since this would lead to unacceptably fast proton decay \cite{Dreiner:1997uz}.
Therefore, we can make the further simplifying assumption that only one or
other of the operators $\hat{L}_i \hat{Q}_j \hat{D}_k$ 
and $\hat{U}_i \hat{D}_j \hat{D}_k$ is present at a time.

When constraining R--violating interactions using experimental data,
it is important to provide a consistent accounting of the corrections
from the R--conserving sector also since they may be sizable depending
on the choice of SUSY parameters.
It is also important to include {\it all} the 
affected observables in a global fit since different observables
may pull the fit values in opposite directions.
This was illustrated in our previous paper \cite{Lebedev:1999vc}
in which the violation of lepton universality was used to constrain
the $\lambda'$ couplings.  There, the $Z$--lineshape observables alone
preferred a 2$\sigma$ limit of $|\lambda'_{33k}| < 0.30$, but a global
fit resulted in $|\lambda'_{33k}| < 0.42$.
Neither of these points were considered in previous works such as
Ref.~\cite{Bhattacharyya:1995bw} where R--conserving corrections were
neglected altogether, and only corrections 
to the ratios of hadronic to leptonic partial widths 
$R_\ell = \Gamma_{\rm had}/\Gamma_{\ell\bar{\ell}}$ ($\ell=e,\mu,\tau$)
were considered. 
It is clear that these ratios receive R--conserving
corrections from top--Higgs and chargino--sfermion loops, as well as
QCD and gluino corrections which depend strongly on $\alpha_s(M_Z)$. 
Therefore, the resulting 1$\sigma$ bound of 
$|\lambda^{\prime \prime}_{3jk}| \leq 0.50$
of Ref.~\cite{Bhattacharyya:1995bw} is hardly robust.

In this paper, we consider all the purely 
hadronic observables which can be expressed as ratios of 
the quark couplings to the $Z$, {\it i.e.} the ratios of 
hadronic partial widths and the parity violating asymmetry parameters.
These are unaffected by QCD and gluino corrections since 
they modify the left-- and right--handed quark couplings multiplicatively, 
leaving the ratios of the couplings intact.\footnote{We assume 
degenerate squark masses.}
The rest of the R--conserving sector induces relevant
corrections to the {\it left--handed} quark couplings only, whereas the
R--breaking sector affects predominantly the 
{\it right--handed} quark couplings.
This allows us to parametrize and constrain 
the R--conserving and R--breaking corrections separately,
thereby constraining the R--breaking sector without making 
{\it ad hoc} assumptions about the R--conserving sector.\footnote{%
Similar methods have been used in Ref.~\cite{DPF:94} and \cite{LOINAZ:99}
to constrain flavor specific vertex corrections while taking
into account the flavor universal oblique corrections.}
Also, since we incorporate into our fit the corrections to the
forward--backward and polarization asymmetries which are much more
sensitive than $R_\ell$ to the shifts in the {\it right--handed} quark 
couplings, we are able to substantially improve the
limits on the R--breaking interactions.

This paper is organized as follows:
In section~\ref{sec:Prelim} we discuss the approximations
we make to simplify our analysis.
In sections~\ref{sec:lambdaprime} and \ref{sec:lambdadoubleprime}
we discuss how the $\lambda'$ and $\lambda''$ interactions
affect the couplings of the quarks to the $Z$.
Section~\ref{sec:Rconserving} discusses the corrections from
the R--conserving sector.
In sections~\ref{sec:Fit} and \ref{sec:Limits}, 
we parametrize the R--conserving and
R--violating corrections to the LEP/SLD observables
and fit them to the latest expermental data, and then translate
the result into limits on $\lambda'$ and $\lambda''$.
In section~\ref{sec:Bayesian}, we provide the Bayesian confidence
limits on $\lambda'$ and $\lambda''$ with the a priori assumption
that the MSSM with R--violation is the correct underlying theory.
Section~\ref{sec:Conclusion} concludes.

%%%%%%%%%%%%%%%%%%%%%%%%%%%%%%%%%%%%%%%%%%%%%%%%%%%%%%%%%%%%%%%%%%%%%%%%%%%%%%
\section{Preliminary Simplifications}
\label{sec:Prelim}

As we stated in the introduction, we only consider supersymmetric
R--violating interactions and neglect the effects of soft--breaking 
R--violating terms.\footnote{See Ref.~\cite{PILAFTSIS:96} for a discussion on 
their possible effects.  LEP/SLD hadronic observables can also be
affected by resonant sneutrino production \cite{Erler:1997ww}.} 
We also neglect the $\lambda$ interactions and
consider only the $\lambda'$ or the $\lambda''$ interactions at a time.
In addition, left--right squark mixing is neglected since 
their effects are expected to be unimportant \cite{Lebedev:1999vc}.
Even with these simplifications, we still have 27 independent
$\lambda'$ couplings or 9 independent $\lambda''$ couplings which
must be considered.

However, a careful look at the diagrams which must be calculated
begets a further simplification.
The corrections to the $Zq\bar{q}$ vertex generated by the
$\lambda'$ and $\lambda''$ interactions in the superpotential 
fall into four classes:  
\begin{enumerate}
\item
%[\raisebox{-1cm}{\begin{picture}(2.5,2.4)(0,0)
%\epsfbox[0 0 90 60]{ssf.ps}
%\end{picture}}]
One particle irreducible (1PI) 
diagrams with two scalars and one fermion in the loop.

\item 
%[\raisebox{-1cm}{\begin{picture}(2.5,2.4)(0,0)
%\epsfbox[0 0 90 60]{ffs.ps}
%\end{picture}}]
1PI diagrams with two fermions and one scalar in the loop.

\item 
%[\raisebox{-1cm}{\begin{picture}(2.5,2.4)(0,0)
%\epsfbox[0 0 90 60]{mms.ps}
%\end{picture}}]
1PI diagrams with two fermions and one scalar in the loop, 
with two mass insertions on the fermion lines.

\item 
%[\raisebox{-1cm}{\begin{picture}(2.5,2.4)(0,0)
%\epsfbox[0 0 90 60]{wfu.ps}
%\end{picture}}]
Fermion wavefunction renormalization diagrams.

\end{enumerate}
In these diagrams,
it is clear that the scalar must be the sparticle while the
internal fermion must be an ordinary lepton or quark. 
The invariant masses of
the external gauge boson and the external fermions must be
set to $m_Z^2$ and $m_q^2 \approx 0$, respectively.

Of the four classes, the third class is finite while the 
$1/\epsilon$ poles of the first two classes cancel against the poles 
in the fermion wavefunction renormalizations.
An explicit evaluation of the finite pieces of the diagrams
reveal \cite{Lebedev:1999vc} that they lead to numerically significant
contributions only when the fermion running in the loop is heavy. 
In fact, the amplitude of a diagram with a massless internal
fermion is only about 10\% of that with an internal top--quark, 
assuming that the scalar (sfermion) mass is the same.
Since each diagram is proportional to $\lambda'$ or $\lambda''$
{\it squared}, dropping these 10\% contributions to the amplitude
will result in a 5\% uncertainty in the limits obtained for
the $\lambda'$ and $\lambda''$.
We can therefore neglect any diagram which does not involve a
top--quark.  
This means that the only R--violating couplings which
are relevant to our discussion are $\lambda'_{i3k}$ (9 parameters) and
$\lambda''_{3jk}$ (3 parameters).

Furthermore, the values of the top--quark diagrams at 
$m_Z^2\rightarrow 0$ provide an excellent approximation to the full integral.
Henceforth we work in this approximation. (We note that the diagrams 
carrying only massless fermions would vanish in this limit 
even if we had previously retained them.) 

In the limit $m_Z^2\rightarrow 0$, 
the four classes of diagrams can be written in terms of the 
$1/\epsilon$ pole piece and two independent functions of 
the fermion--scalar mass ratio $x = m_f^2/m_s^2$
which we call $f(x)$ and $g(x)$.  
Their explicit forms are shown in the Appendix.
We note that $g(x)$, the function which appears 
in the two--scalar one--fermion,
and the fermion wavefunction renormalization diagrams, 
vanishes rapidly as $x\rightarrow 1$.
Thus, the finite pieces of the the two--scalar one--fermion and 
the wavefunction renormalization diagrams may be neglected 
for small scalar--fermion mass splittings.  
Note further that if the poles of these diagrams cancel
(as is the case for diagrams involving gluinos, for example), 
the $m_Z^2=0$ finite pieces will also cancel.
These considerations apply equally to R--conserving corrections
leading to significant simplifications in their contributions
as well, the details of which will be discussed in 
Sec.~\ref{sec:Rconserving}.

%%%%%%%%%%%%%%%%%%%%%%%%%%%%%%%%%%%%%%%%%%%%%%%%%%%%%%%%%%%%%%%%%%%%%%%%%%%%%%
\section{Corrections from the $\lambda^\prime$ Interactions}
\label{sec:lambdaprime}

The R--parity violating $\lambda^\prime$ interactions expressed in terms of 
the component fields take the form
\begin{eqnarray}
\lefteqn{\Delta{\cal L}'_{\not{R}}
    = \lambda^{\prime}_{ijk}
      \bigg[ \tilde{\nu}_{iL} \overline{d}_{kR} d_{jL}
            + \tilde{d}_{jL} \overline{d}_{kR} \nu_{iL}
            + \tilde{d}^*_{kR} \overline{\nu}^c_{iL} d_{jL}}
\cr
& &         - (\tilde{e}_{iL} \overline{d}_{kR} u_{jL}
            + \tilde{u}_{jL} \overline{d}_{kR} e_{iL}
            + \tilde{d}^*_{kR} \overline{e}^c_{iL} u_{jL})
      \bigg]\; + \;{h.c.}
\label{eq:Lagrangian1}
\end{eqnarray}
As discussed in the previous section,
the dominant corrections to hadronic $Z$ decays 
from these interactions are those which involve the top--quark.
These are shown in Fig.~\ref{Zdecaylambdaprime}.
(The diagrams with an internal top--quark and external leptons
were considered in Ref.~\cite{Lebedev:1999vc}.)
This necessarily means that only the
couplings of the {\it right--handed down--type} quarks $d_{i_R}$ to the $Z$
are corrected in our approximation.  

Using notation established in Ref.~\cite{Lebedev:1999vc}, 
the corrections to the $Z$ decay amplitude from these diagrams are
\begin{eqnarray}
\lefteqn{%
- | \lambda^\prime_{i3k} |^2
\Biggl[ -i\frac{ g }{ \cos\theta_W }\,
        Z^\mu(p+q)\,\bar{q}_{k_R}(p) \gamma_\mu q_{k_R}(q)\,
\Biggr] \times}
\qquad & & \cr
(1a) & : & 2 h_{e_L}
     \hat{C}_{24}\left( 0,0,m_Z^2;m_{t},m_{\tilde{e}_{iL}},m_{\tilde{e}_{iL}}
                 \right)
\cr
(1b) & : & \phantom{2}h_{u_L}
\left[ (d-2)\hat{C}_{24}\left( 0,0,m_Z^2;m_{\tilde{e}_{iL}},m_{t},m_{t}
                        \right) 
\right. \cr
     &   &
\left.\qquad\quad
    - m_Z^2 \hat{C}_{23}\left( 0,0,m_Z^2;m_{\tilde{e}_{iL}},m_{t},m_{t}
                        \right) 
\right]
\cr
(1c) & : &  - h_{u_R} m_{t}^2  
           \hat{C}_0\left( 0,0,m_Z^2;m_{\tilde{e}_{iL}},m_{t},m_{t}
                    \right)
\cr
(1d) & + & (1e)\;:\; 2 h_{d_R} 
                  B_1\left( 0;m_{t},m_{\tilde{e}_{iL}} \right)
\end{eqnarray}
where
\begin{equation}
h_{f_L} = I_{3f} - Q_f \sin^2\theta_W, \qquad
h_{f_R} =        - Q_f \sin^2\theta_W.
\end{equation}
The tree level amplitude is $h_{d_{iR}}$ times the expression in the 
square brackets.
These corrections can be expressed as a shift in the coupling $h_{d_{iR}}$:
\widetext
\begin{eqnarray}
\delta h_{i3k}
& \equiv & -  |\lambda^{\prime}_{i3k}|^2
\bigg[ 2 h_{e_L}
        \hat{C}_{24}\left( m_{t},m_{\tilde{e}_{iL}},m_{\tilde{e}_{iL}}
                    \right) \cr
&   &\qquad\qquad
       + h_{u_L}
        \left\{ (d-2)\hat{C}_{24}\left(m_{\tilde{e}_{iL}},m_{t},m_{t}
                                \right) 
            - m_Z^2 \hat{C}_{23}\left(m_{\tilde{e}_{iL}},m_{t},m_{t}
                                \right) 
        \right\}  \cr 
&   &\qquad\qquad
       - h_{u_R} m_{t}^2  
        \hat{C}_0\left(m_{\tilde{e}_{iL}},m_{t},m_{t}
                 \right) \cr
&   &\qquad\qquad
       + h_{d_R} B_1\left( m_{t},m_{\tilde{e}_{iL}}
                    \right).
\bigg]
\end{eqnarray}
Henceforth we assume a common slepton mass 
$m_{\tilde{e}_{iL}} = m_{\tilde{e}}$, $i=1,2,3$.
The full expression for $\delta h_{i3k}$
is well approximated by the leading $m_Z^2 = 0$ piece of the 
expansion in the $Z$ mass:
\begin{equation}
\delta h_{i3k}
\approx  \frac{1}{ 2(4\pi)^2 } | \lambda^{\prime}_{i3k} |^2 F(x)
\end{equation}
where  
\begin{equation}
F(x) = f(x)+g(x) = \frac{x}{1-x} \left( 1+ \frac{ 1 }{ 1-x }\ln x
                                 \right),
\qquad
x = \frac{m_t^2}{m_{\tilde{e}}^2}.
\label{eq:F}
\end{equation}
For $m_{\tilde{e}}=100 {\rm GeV}$, this becomes
\begin{equation}
\delta h_{i3k}
\approx -0.215 \%\, | \lambda^{\prime}_{i3k} |^2.
\label{eq:hlambdaprime}
\end{equation}
The full shift to the coupling of the quark $d_{k_R}$ to the $Z$ 
due to R--violating $\lambda^{\prime}$ interactions
is then obtained by summing over the slepton generation index $i$
\begin{eqnarray}
\delta h_{d_k}^{\not R}
& = & \sum_{i}\delta h_{i3k} \cr
& \approx & -0.215\% \sum_i |\lambda^{\prime}_{i3k}|^2
\label{eq:deltahprime}
\end{eqnarray} 
Observe that this is a different combination of $\lambda^{\prime}$ 
couplings than the combination $\sum_k |\lambda^{\prime}_{i3k}|^2$ 
which is constrained by lepton universality in Ref.~\cite{Lebedev:1999vc}.

%There are also R--violating $\lambda^{\prime}$ shifts to the $u_{i_R}$ and 
%$\nu_{i_L}$ couplings to $Z$.  
%However, such shifts are generated only by diagrams with light quarks
%in the loops.  
%These contributions will be numerically small and are neglected 
%in our approximation.
%Thus, all significant R--violating $\lambda^{\prime}$
%shifts to $Z$--pole observables appear as shifts in the
%coupling of the $Z$ to left--handed charged leptons 
%(which we considered in Ref.~\cite{Lebedev:1999vc})
%or to right--handed down--type quarks, which we consider here.

%%%%%%%%%%%%%%%%%%%%%%%%%%%%%%%%%%%%%%%%%%%%%%%%%%%%%%%%%%%%%%%%%%%%%%%%%%%%%%
\section{Corrections from the $\lambda^{\prime \prime}$ Interactions}
\label{sec:lambdadoubleprime}

The R--parity violating $\lambda^{\prime\prime}$ interactions expressed 
in terms of the component fields take the form
\begin{equation}
\Delta{\cal L}''_{\not{R}} 
= \frac{1}{2} \lambda''_{ijk}
\bigg[ u_{i}^{c} d_{j}^{c} \tilde d^{*}_{k}
     + u_{i}^{c} \tilde d_{j}^{*} d^{c}_{k}
     + \tilde u_{i}^{*} d_{j}^{c} d^{c}_{k} \bigg]\; + \;{h.c.}
\label{eq:Lagrangian2}
\end{equation}
The $SU(3)$ color indices are suppressed.
Note that $\lambda^{\prime \prime}_{ijk}$ is
antisymmetric in the last two indices due to color anti--symmetrization. 

Again, the corrections involving a top--quark are necessarily 
those with a right-handed down--type quark on the external legs
as shown in Fig.~\ref{Zdecaylambdadoubleprime}.
Their respective contributions to the amplitude are:
\begin{eqnarray}
\lefteqn{%
-2 | \lambda^{\prime \prime}_{3jk} |^2
\Biggl[ -i\frac{ g }{ \cos\theta_W }\,
        Z^\mu(p+q)\,\bar{q}_{jR}(p) \gamma_\mu q_{jR}(q)\,
\Biggr] \times}
\qquad & & 
\cr
(2a) & : & - 2 h_{d_R}
     \hat{C}_{24}\left( 0,0,m_Z^2;m_{t},m_{\tilde{d}_{kR}},m_{\tilde{d}_{kR}}
                 \right)
\cr
(2b) & : &   - h_{u_R}
\left[ (d-2)\hat{C}_{24}\left( 0,0,m_Z^2;m_{\tilde{d}_{kR}},m_{t},m_{t}
                        \right) 
\right. \cr
     &   &
\left.\qquad\quad
    - m_Z^2 \hat{C}_{23}\left( 0,0,m_Z^2;m_{\tilde{d}_{kR}},m_{t},m_{t}
                        \right) 
\right]
\cr
(2c) & : &   h_{u_L} m_{t}^2  
           \hat{C}_0\left( 0,0,m_Z^2;m_{\tilde{d}_{kR}},m_{t},m_{t}
                    \right)
\cr
(2d) & + & (2e)\;:\; 2 h_{d_R} 
                  B_1\left( 0;m_{t},m_{\tilde{d}_{kR}} \right)
\end{eqnarray}
The common leading factor of 2 in this equation is a consequence of the
identity
\[
\varepsilon_{abc}\varepsilon_{a'bc} = 2\delta_{aa'}.
\]
These corrections shift the coupling of the right--handed
down--type quark $h_{d_{j_R}}$ to the $Z$ by
\widetext
\begin{eqnarray}
\delta h_{3jk}
& \equiv & - 2 |\lambda^{\prime \prime}_{3jk}|^2
\bigg[ -2 h_{d_R}
        \hat{C}_{24}\left( m_{t},m_{\tilde{d}_{kR}},m_{\tilde{d}_{kR}}
                    \right) \cr
&   &\qquad\qquad
       - h_{u_R}
        \left\{ (d-2)\hat{C}_{24}\left(m_{\tilde{d}_{kR}},m_{t},m_{t}
                                \right) 
            - m_Z^2 \hat{C}_{23}\left(m_{\tilde{d}_{kR}},m_{t},m_{t}
                                \right) 
        \right\}  \cr 
&   &\qquad\qquad
       + h_{u_L} m_{t}^2  
        \hat{C}_0\left(m_{\tilde{d}_{kR}},m_{t},m_{t}
                 \right) \cr
&   &\qquad\qquad
       + h_{d_R} B_1\left( m_{t},m_{\tilde{d}_{kR}}
                    \right)
\bigg]
\end{eqnarray}
We assume a common squark mass 
$m_{\tilde{d}_{kR}} = m_{\tilde{d}}$, $k=1,2,3$.
As in the $\lambda^\prime$ case, we have neglected all diagrams which 
vanish in the limit $m_Z \rightarrow 0$ in the above expression.
Applying the same approximation to the leading diagrams leaves:
\begin{equation}
\delta h_{3jk}
\approx  \frac{1}{ (4\pi)^2 } | \lambda^{\prime \prime}_{3jk} |^2 F(x)
\end{equation}
with $x = m_t^2/m_{\tilde{d}}^2$.
For $m_{\tilde{d}}=100 {\rm GeV}$, this becomes
\begin{equation}
\delta h_{3jk}
\approx -0.43 \%\, | \lambda^{\prime \prime}_{3jk} |^2.
\label{eq:hlambdadoubleprime}
\end{equation}
The full shift to the coupling of the quark $d_{j_R}$ to the $Z$ 
due to R--violating $\lambda^{\prime \prime}$ interactions
is then obtained by summing over the slepton generation index $k$
\begin{eqnarray}
\delta h_{d_{jR}}^{\not R}
& = & \sum_{k}\delta h_{3jk} \cr
& \approx & -0.43\% \sum_k |\lambda^{\prime \prime}_{3jk}|^2.
\label{eq:deltahdoubleprime}
\end{eqnarray} 
%

%There are also R--violating $\lambda^{\prime \prime}$ contributions to the 
%$u_{i_R}$ coupling to the $Z$.  
%However, as in the $\lambda^{\prime}$ case the corresponding diagrams involve 
%only light quarks in loops and are thus neglected.

In contrast to the $\lambda^{\prime}$ case, $\lambda^{\prime \prime}$ 
interactions do {\it not} correct any of the lepton couplings to the $Z$.  
Thus they do not give rise to lepton universality violations, 
and no additional constraints on
$\lambda^{\prime \prime}$ couplings arise from analysis of the lepton sector.
Thus, all significant R--violating $\lambda^{\prime \prime}$ shifts to $Z$ 
pole observables appear as shifts to the effective coupling of the $Z$ to 
right--handed down--type quarks.

%%%%%%%%%%%%%%%%%%%%%%%%%%%%%%%%%%%%%%%%%%%%%%%%%%%%%%%%%%%%%%%%%%%%%%%%%%%%%%
\section{Corrections from R--Conserving Interactions}
\label{sec:Rconserving}

As stressed in the introduction, in order to isolate the effects of
R--violating interactions we must properly parametrize 
the R--conserving radiative corrections (a partial study of 
these effects has also been performed in \cite{Yang:1999ms}).  
We work in the limit of degenerate sfermion masses and 
$\tan\beta$ not large.
In this limit, only two parameters are necessary to
account for R--conserving effects.

We list all relevant vertex corrections from R--conserving MSSM
interactions:
\begin{description}

\item [chargino--sfermion loops :]\hfill\\
The fermion interactions with the gaugino component of the chargino can
generate substantial corrections to the left--handed couplings of all the 
fermions (Fig.~\ref{CharginoSfermion}).  
The correction 
to the up--type and down--type quark couplings from the diagrams 
shown in Figs.~\ref{CharginoSfermion}b,c,d are proportional to
\begin{eqnarray}
\delta h_{u_L}^{(\ref{CharginoSfermion}b,c,d)}
& \propto & 2h_{d_L} 
\,\hat{C}_{24}(m_{\tilde{\chi}},m_{\tilde{d}_L},m_{\tilde{d}_L})
+ h_{u_L} B_1(m_{\tilde{\chi}},m_{\tilde{d}_L}), \cr
\delta h_{d_L}^{(\ref{CharginoSfermion}b,c,d)}
& \propto & 2h_{u_L} 
\,\hat{C}_{24}(m_{\tilde{\chi}},m_{\tilde{u}_L},m_{\tilde{u}_L})
+ h_{d_L} B_1(m_{\tilde{\chi}},m_{\tilde{u}_L}),
\end{eqnarray}
where the dependence on the external momenta have been suppressed.
In the limit $m_Z^2\rightarrow 0$ and $m_{\tilde{u}_L} = m_{\tilde{d}_L}$,
it is clear (see Appendix) that
\begin{equation}
\delta h_{u_L}^{(\ref{CharginoSfermion}b,c,d)} = 
- \delta h_{d_L}^{(\ref{CharginoSfermion}b,c,d)}
\propto (h_{u_L}-h_{d_L}) = (1 -\sin^2\theta_W).
\end{equation}
A similar relation 
exists for the correction to the leptonic vertices provided
$m_{\tilde{\nu}_L} = m_{\tilde{e}_L}$.
In addition, the diagram of Fig.~\ref{CharginoSfermion}a 
changes sign with the isospin of the final--state fermion.
As a result, the combined contribution from all the diagrams 
in Fig.~\ref{CharginoSfermion} is proportional to the isospin of the 
final--state fermion but otherwise universal in the limit that all the 
(left--handed) squark and slepton masses are degenerate.  

A shift proportional to the isospin can be written as an overall 
multiplicative change in the coupling and a shift in
the effective value of $\sin^2{\theta_W}$:
\begin{eqnarray}
h_{f_L}
& = & I_{3f} ( 1+\delta ) - Q\sin^2\theta_W 
\;=\; ( 1+\delta )
      \left( I_{3f}  - Q\,\frac{ \sin^2\theta_W }{ 1+\delta }
      \right)\cr
h_{f_R}
& = & \phantom{ I_{3f} ( 1+\delta ) } - Q\sin^2\theta_W
\;=\; ( 1+\delta )
      \left( \phantom{ I_{3f} } - Q\,\frac{ \sin^2\theta_W }{ 1+\delta }
      \right)
\end{eqnarray}
Since we utilize only observables which are ratios of couplings, 
the multiplicative correction cancels and only the shift 
in $\sin^2\theta_W$ is measurable.

\item[charged Higgs--top (Higgsino--stop) loops :]\hfill\\
The charged Higgs--top corrections
(Figs.~\ref{TopHiggsLoop}a,d,e)
and the supersymmetrized versions of these diagrams 
containing the chargino--right handed stop loops 
(Figs.~\ref{TopHiggsLoop}b,c),
as well as the corresponding wavefunction renormalizations
are peculiar to the $b_{L}$ final states.
Higgs and higgsino couplings to the $b_{R}$ and other 
fermion final states are suppressed by
the light fermion masses and are neglected.
The model generically predicts $\dhbLH > 0$.  
This is a result of the fact that (1) the charged Higgs contribution
is always positive \cite{Haber:1999zh} and 
(2) explicit calculation shows that 
the leading $p^2 / m^2$ Higgsino contribution
vanishes.{\footnote{For the purpose of this analysis, we treat 
Higgsino and gaugino contributions separately.}}

\item[gluino--squark loops :]\hfill\\
The leading ($m_Z^2=0$) contributions from the diagrams 
shown in Fig.~\ref{GluinoSquark} vanish as a 
result of the relation (see Appendix)
\begin{equation}
\biggl[\, 2\,\hat{C}_{24}(0,0,m_Z^2;m_{\tilde{g}},m_{\tilde{q}},m_{\tilde{q}})
         + B_1(0;m_{\tilde{g}},m_{\tilde{q}})\,
\biggr]_{m_Z^2=0} = 0.
\label{eq:gluino}
\end{equation}
Even for the subleading terms, in the limit of degenerate squark masses 
the gluino--squark loops induce only a universal shift to all of the 
$Z$--quark couplings.
Just as for QCD corrections, this shift cancels in the
ratios of hadronic widths and asymmetries and
therefore does not enter into our analysis.

\item [neutralino--sfermion loops :]\hfill\\
The gaugino component of the neutralino generates shifts 
to all of the couplings (Fig.~\ref{Neutralino}).  
However, the only (potentially) significant diagrams are all 
of the two scalar, three sparticle variety.  
When these are combined with wavefunction renormalization
diagrams, the leading $m_Z^2=0$ pieces cancel in the sum
(see Eq.~\ref{eq:gluino}). Therefore, these can be neglected altogether.

\item [oblique corrections :]\hfill\\
In addition to all these vertex corrections, R--conserving
interactions can also affect $Z$--peak observables through vacuum polarization
diagrams, aka the oblique corrections.
Oblique corrections can all be subsumed into a shift in the
$\rho$ parameter and the effective value of $\sin^2\theta_W$, the
first of which cancels in all of the observables that we consider
\cite{PESKIN:90}.

\end{description}
Thus, in our approximation, the only
R--conserving effects we need to consider are
(1) a shift in the left--handed $b$ coupling from charged Higgs/Higgsino 
corrections, and (2) a universal shift in the effective value of
$\sin^2\theta_W$ which subsumes the isospin--proportional correction due to
chargino--squark loops as well as the oblique corrections.

%%%%%%%%%%%%%%%%%%%%%%%%%%%%%%%%%%%%%%%%%%%%%%%%%%%%%%%%%%%%%%%%%%%%%%%%%%%%%%
\section{Fit to the data}
\label{sec:Fit}

As we have seen, 
the R--violating $\lambda^{\prime}$ couplings correct
the left--handed couplings of the charged leptons and the
right--handed couplings of the down--type quarks while the
$\lambda^{\prime \prime}$ couplings correct the right--handed couplings of the
down--type quarks only.
The R--conserving chargino--sfermion correction is absorbed into
a universal shift of $\sin^2{\theta_W}$ (provided that all the
sfermions are degenerate) while the Higgs--top correction is only
relevant for the left--handed $b$ quark.
Since we have already discussed the limits placed on 
the $\lambda^{\prime}$ couplings from the leptonic observables 
in a previous paper \cite{Lebedev:1999vc}, we will concentrate on
the corrections to the quark observables and perform
a fit which encompassed both the $\lambda^{\prime}$ and $\lambda^{\prime \prime}$ cases.

In order to constrain the size of these corrections
we will use the \textit{ratios} of the hadronic parital widths
\begin{eqnarray*}
R_q  & = & \frac{ \Gamma_{q\bar{q}} }
                { \Gamma_{\rm had}  }
     \;=\; \dfrac{ h_{q_L}^2 + h_{q_R}^2 }
                { \sum_{q'=u,d,s,c,b} (h_{q'_L}^2 + h_{q'_R}^2) 
                },\qquad (q=c,b)\cr
R'_q & = & \frac{ \Gamma_{q\bar{q}} }
                { \Gamma_{u\bar{u}}
                + \Gamma_{d\bar{d}}
                + \Gamma_{s\bar{s}} }
     \;=\; \dfrac{ h_{q_L}^2 + h_{q_R}^2 }
                { \sum_{q=u,d,s} (h_{q'_L}^2 + h_{q'_R}^2) 
                },\qquad (q=u,d,s)
\end{eqnarray*}
and the parity--violating asymmetry parameters 
\[
A_q = \frac{ h_{q_L}^2 - h_{q_R}^2 }
           { h_{q_L}^2 + h_{q_R}^2 },\qquad
A_{\rm FB}(q) = \frac{3}{4}A_e A_q,\qquad (q=u,d,s,c,b).
\]
These observables have the convenient property that
(1) they are insensitive to QCD and gluino--squark corrections, and
(2) the only dependence on oblique corrections (vacuum polarizations)
comes from a shift in the effective value of $\sin^2\theta_W$.
This will permit us to constrain the parameters we are interested in
without complicating the fit procedure by introducing 
gluon/gluino corrections or corrections to the $\rho$
parameter.

Of the leptonic observables, we will include the ratio of electron 
to neutrino widths $R_{\nu/e} = \Gamma_{\nu\bar{\nu}}/\Gamma_{e^+e^-}$
and the electron asymmetry parameters $A_e$ and
$A_{\rm FB}(e) = \frac{3}{4}A_e^2$ 
to help constrain the universal R--conserving and 
oblique corrections.  
Corrections to $A_e$ must be considered
in any case since it is present in the hadronic observables $A_{\rm FB}(q)$.
Though the left--handed lepton couplings
receive corrections from the $\lambda^{\prime}$ interactions, the
size of the correction particular to the electron is already so tightly
constrained to be small by other experiments that we can neglect it entirely.  
We drop all $\mu$ or $\tau$ dependent
observables from our fit so that we can use the result
to constrain both the $\lambda^{\prime}$ and $\lambda^{\prime \prime}$ cases.

In table~\ref{LEP-SLD-DATA} we list the experimental data
we use in our fit with correlation matrices shown in
tables~\ref{lineshape-correlations} and \ref{heavy-correlations}.
We caution the reader that many of these numbers are preliminary results
announced during the summer 1999 conferences so they, and our resulting
fit derived from them, may be subject to change.
Some comments are in order:
\begin{enumerate}
\item
The ratio 
\[
R_{\nu/e}
= \frac{ \Gamma_{\nu\bar{\nu}} }
       { \Gamma_{e^+e^-} }
= \frac{ h_{\nu_L}^2 }
       { h_{e_L}^2 + h_{e_R}^2 }
\]
is calculated from the first six $Z$--lineshape observables.
Its correlation to $A_{\rm FB}(e)$ is $+28\%$.  Its correlations to
the $\mu$ and $\tau$ observables, which we drop, are negligibly small.

\item
The $\tau$ polarization data has been updated from Ref.~\cite{LEP:98}
with new numbers from DELPHI \cite{MNICH:99,SLD:99}.
We keep only $A_e$ and drop $A_\tau$.

\item
The SLD value of $A_{\rm LR}$ (which is the same thing as
$A_e$) is from hadronic events only.  $A_e$ is from the leptonic
events.   Its correlations to the dropped $A_\mu$ and $A_\tau$
are negligibly weak. 
(The errors are dominated by statistics \cite{ROWSON:99}.)

\item
The OPAL measurements of $R'_s$, $A_{\rm FB}(s)$, and $A_{\rm FB}(u)$
assume Standard Model values of $R'_d = 0.359$ and $A_{\rm FB}(d) = 0.100$.
To account for the shifts in the down observables in our model, the
data should be interpreted as constraining the following linear combinations:
\begin{eqnarray*}
R^{\prime *}_s
& = & R^{\prime}_s + 1.83\;[\; R'_d - 0.359 \;]  \cr
A^*_{\rm FB}(s)
& = & A_{\rm FB}(s) - 0.32\;[\; A_{\rm FB}(d) - 0.100 \;] \cr
A^*_{\rm FB}(u)
& = & A_{\rm FB}(u) - 1.42\;[\; A_{\rm FB}(d) - 0.100 \;]
\end{eqnarray*}
There is a $+31\%$ correlation between $A^*_{\rm FB}(s)$ and
$A^*_{\rm FB}(u)$ \cite{OPAL:97}.

\item
The DELPHI measurement of $A_{\rm FB}(s)$ assumes
standard model values of $A_{\rm FB}(u) = 0.0736$ and
$A_{\rm FB}(d) = 0.1031$.  It should be interpreted as a measurement
of the following linear combination \cite{DELPHI:99}:
\[
A^{**}_{\rm FB}(s)
= A_{\rm FB}(s) - 0.156 \;[\; A_{\rm FB}(u) - 0.0736 \;]
                - 0.117 \;[\; A_{\rm FB}(d) - 0.1031 \;].
\]

\item
The SLD measurement of $A_s$ \cite{As-SLD:99} assumes standard model
values for $A_u$, $A_d$, $R'_u$, and $R'_d$.  
(The dependence on the heavy flavor variables are weak and negligible.)
To account for shifts in these input parameters the
measurement should be intepreted as constraining \cite{MULLER:99}
\begin{eqnarray*}
A_s^{*}
& = & A_s - 0.0602 \;[\; A_u - 0.668 \;]
          - 0.0467 \;[\; A_d - 0.936 \;] \cr
&   &\phantom{A_s}
          - 1.32 \;[\; R'_u - 0.280 \;]
          - 1.20  \;[\; R'_d - 0.360 \;]
\end{eqnarray*}

\item
The heavy flavor data is the combined fit to the LEP and SLD data
compiled by the LEP Electroweak Working Group \cite{MNICH:99}.
The central values of $A_b$ and $A_c$ are shifted compared to the
original SLD values of
\begin{eqnarray*}
A_b & = & 0.905 \pm 0.026 \cr
A_c & = & 0.634 \pm 0.027
\end{eqnarray*}
Using these numbers instead of those shown in Table~\ref{LEP-SLD-DATA}
will result in a slightly tighter constraint on the R--violating couplings,
but we will present the results using the LEPEWWG numbers to be
on the conservative side.

\end{enumerate}

We denote the shift in $\sin^2\theta_W$ due to oblique and
chargino--sfermion corrections by $\dss$, 
and the shift from the Higgs interactions
specific to the left--handed coupling of the $b$ by $\dhbLH$.
The R--violating shifts specific to the right--handed couplings
of the $d$, $s$, and $b$ quarks are denoted
$\dhdR$, $\dhsR$, and $\dhbR$.
Then the shifts in the couplings of the quarks, the electron, and
the neutrino are given by:
\begin{eqnarray*}
\delta h_{\nu_L} & = & \phantom{-}0        \cr
\delta h_{e_L}   & = & \phantom{-}\dss  \cr
\delta h_{e_R}   & = & \phantom{-}\dss        \cr
\delta h_{u_L}   & = & - \frac{2}{3}\dss   \cr
\delta h_{u_R}   & = & - \frac{2}{3}\dss         \cr
\delta h_{d_L}   & = & \phantom{-}\frac{1}{3}\dss   \cr
\delta h_{d_R}   & = & \phantom{-}\frac{1}{3}\dss + \dhdR \cr
\delta h_{c_L}   & = & - \frac{2}{3}\dss   \cr
\delta h_{c_R}   & = & - \frac{2}{3}\dss         \cr
\delta h_{s_L}   & = & \phantom{-}\frac{1}{3}\dss   \cr
\delta h_{s_R}   & = & \phantom{-}\frac{1}{3}\dss + \dhsR \cr
\delta h_{b_L}   & = & - \frac{1}{3}\dss  + \dhbLH \cr
\delta h_{b_R}   & = & - \frac{1}{3}\dss + \dhbR
\end{eqnarray*}
The dependence of the observables on these fit parameters
can be calculated in a straightforward manner.
For instance, we find:
\begin{eqnarray*}
\frac{ \delta R_{\nu/e} }
     { R_{\nu/e} }
& = & \frac{ 2\,\delta h_{\nu_L} }{ h_{\nu_L} }
    - \frac{ 2 h_{e_L}\delta h_{e_L} + 2 h_{e_R}\delta h_{e_R} }
           { h_{e_L}^2 + h_{e_R}^2 }  \cr
& = & - \left( \frac{ 2 h_{e_L} + 2 h_{e_R} }
                  { h_{e_L}^2 + h_{e_R}^2 }
      \right) \dss \cr
& = & 0.64\,\dss 
\end{eqnarray*}
or
\[
\delta R_{\nu/e} = 1.17\,\dss 
\]
where the coefficient has been calculated assuming $\sin^2\theta_W = 0.2315$.
Similarly,
\begin{eqnarray}
\delta A_e             & = & -7.61\,\dss   \cr
\delta A_{\rm FB}(e)   & = & -1.63\,\dss  \cr
\delta R_s^{\prime *}  & = & \phantom{-}0.151\,\dss 
                           + 0.242\,\dhdR - 0.0058\,\dhsR \cr
\delta A_{\rm FB}^*(u) & = & \phantom{-}3.74\,\dss  + 0.262\,\dhdR \cr
\delta A_{\rm FB}^*(s) & = & -3.72\,\dss 
                           + 0.0558\,\dhdR - 0.174\,\dhsR \cr
\delta A_{\rm FB}^{**}(s)
                       & = & -4.15\,\dss 
                           + 0.020\,\dhdR - 0.174\,\dhsR \cr
\delta A_s^{*} & = & -0.321\,\dss  - 0.0444\,\dhdR - 1.37\,\dhsR \cr
\delta R_b & = & \phantom{-}
                  0.0392\,\dss  - 0.0396\,\dhdR - 0.0396\,\dhsR
                + 0.141\,\dhbR - 0.771\,\dhbLH \cr
\delta R_c & = & -0.0605\,\dss  - 0.0316\,\dhdR - 0.0316\,\dhsR
                - 0.0316\,\dhbR + 0.173\,\dhbLH \cr
\delta A_{\rm FB}(b) & = & -5.40\,\dss 
                           - 0.172\,\dhbR - 0.0315\,\dhbLH \cr
\delta A_{\rm FB}(c) & = & -4.17\,\dss  \cr
\delta A_b           & = & -0.636\,\dss 
                           -1.61\,\dhbR - 0.295\,\dhbLH \cr
\delta A_c           & = & -3.45\,\dss 
\label{eq:FITCOEFFS} 
\end{eqnarray}
Fitting these expressions to the table~\ref{LEP-SLD-DATA} data,
we obtain:
\begin{eqnarray}
\dss   & = & -0.00092 \pm 0.00022 \cr
\dhdR  & = & \phantom{-}0.081 \pm 0.077 \cr
\dhsR  & = & \phantom{-}0.055 \pm 0.043 \cr
\dhbR  & = & \phantom{-}0.026 \pm 0.010 \cr
\dhbLH & = & -0.0031 \pm 0.0042
\label{eq:FITRESULTS}
\end{eqnarray}
with the correlation matrix shown in table~\ref{fit-correlation}.
The quality of the fit was $\chi^2 = 12.0/(16-5)$.  The standard model
predictions were obtained using ZFITTER v.6.21 \cite{ZFITTER:99} using 
$m_t=174.3$~GeV \cite{TOPMASS:99} 
and $m_h=300$~GeV.  Except for $\dss$,
the best--fit values and uncertainties of the parameters are 
virtually unchanged when the Standard Model Higgs mass is varied between 
100~GeV and 1~TeV.
By far the largest contribution to the $\chi^2$ is from those 
observables ($R_{\nu/e}, A_{\rm FB}(c)$ and $A_c$ contribute a 
combined $8.6$) which serve only to 
compete with $A_{\rm LR}$ in determining $\dss$.

In Figs.~\ref{FIG12} through \ref{FIG45},
we show the limits placed on the five parameters by various observables
projected onto two dimensional planes.
Figs.~\ref{FIG12}, \ref{FIG23}, \ref{FIG24}, and \ref{FIG25} show
that the most stringent constraint on $\dhdR$ comes from $A_{FB}^*(u)$,
while Figs.~\ref{FIG13}, \ref{FIG23}, \ref{FIG34}, and \ref{FIG35}
show that the $\dhsR$ is constrained by $A_s^*$ and $A_{\rm FB}^{**}(s)$.  
Figs.~\ref{FIG15}, \ref{FIG25}, \ref{FIG35}, and \ref{FIG45} 
show that $\dhbLH$ is largely fixed by $R_b$.
It is clear from figs.~\ref{FIG24}, \ref{FIG34}, and \ref{FIG45}
that the strongest constraint on $\dhbR$ comes from $A_b$ and $A_{\rm FB}(b)$.
However, a careful look at Fig.~\ref{FIG14} shows that the limit
on $\dhbR$ is strongly correlated with the value of $\dss$.
Because $A_{\rm LR}$ and other measurements prefer a slightly negative
$\dss$, the preferred value of $\dhbR$ from $A_{\rm FB}(b)$
is shifted to the positive side \cite{DPF:94}.

%%%%%%%%%%%%%%%%%%%%%%%%%%%%%%%%%%%%%%%%%%%%%%%%%%%%%%%%%%%%%%%%%%%%%%%%%%%%%%
\section{Limits on $\lambda^{\prime}$ and $\lambda^{\prime \prime}$}
\label{sec:Limits}

Using Eq.~\ref{eq:hlambdaprime}, we can translate our 
fit results in Eq.~\ref{eq:FITRESULTS},
to limits on the $\lambda'$ couplings constants:
\begin{eqnarray}
\sum_i |\lambda^{\prime}_{i31}|^2 & = & -38\phantom{.1} \pm 36 \cr
\sum_i |\lambda^{\prime}_{i32}|^2 & = & -26\phantom{.1} \pm 20 \cr
\sum_i |\lambda^{\prime}_{i33}|^2 & = & -12.1 \pm \phantom{0}4.7.
\end{eqnarray}
The correlations between the fit values of the couplings are relatively small 
(see table~\ref{fit-correlation}).
The 1$\sigma$ (2$\sigma$) [3$\sigma$] upper bounds are then 
\begin{eqnarray}
\sum_i |\lambda^{\prime}_{i31}|^2 
& \leq & -2\phantom{.4}\;(34)\;[70] \cr
\sum_i |\lambda^{\prime}_{i32}|^2 
& \leq & -6\phantom{.4}\;(14)\;[34] \cr
\sum_i |\lambda^{\prime}_{i33}|^2 
& \leq & -7.4 \;(-2.8)\;[1.9]
\end{eqnarray}
This imposes the following 
(2$\sigma$) [3$\sigma$] limits on the individual couplings
in the sum:
\begin{eqnarray}
|\lambda^{\prime}_{i31}| & \leq &( 5.8 )\;[ 8.4 ] \cr
|\lambda^{\prime}_{i32}| & \leq &( 3.8 )\;[ 5.9 ] \cr
|\lambda^{\prime}_{i33}| & \leq &( \phantom{3.8} ) \;[ 1.4 ] 
\end{eqnarray}
For $i=1$ and $i=2$, stronger constraints at the $2 \sigma$ level
on the relevant couplings 
are available from other types of experiments \cite{Allanach:1999ic}, 
so these constraints fail to improve previous results.
The strongest constraint is on $i=3$, where 
the best--fit value of the sum of squared couplings is 
negative even at $2 \sigma$.
This constitutes a significant improvement in the upper bound on 
$|\lambda^{\prime}_{i33}|$ over previous bounds on these couplings
which were nonzero at the $2 \sigma$ level.
These results are complementary to those obtained in 
Ref.~\cite{Lebedev:1999vc}, in 
which a different combination of $\lambda^{\prime}$ couplings was constrained.
In particular, for the $\lambda^{\prime}_{i33}$ couplings, the $\sigma$ 
from the lepton universality constraints is much smaller, 
but the best--fit value of the squared couplings from the hadronic 
constraint is negative by an even greater statistical significance.\footnote{%
A strong independent constraint on $\lambda'_{i33}$ will be available
from the measurement of the invisible width of $\Upsilon$ resonance
\cite{CHANG:98}.}

Next, we consider the constraints on the $\lambda''$ couplings.
These couplings have hitherto been constrained by experiment only 
weakly or not at all.  
Using Eq.~\ref{eq:hlambdadoubleprime}, we can translate 
Eq.~\ref{eq:FITRESULTS} into the bounds:
\begin{eqnarray}
\sum_k |\lambda^{\prime \prime}_{31k}|^2 & = & -19 \pm 18 \cr
\sum_k |\lambda^{\prime \prime}_{32k}|^2 & = & -13 \pm 10 \cr
\sum_k |\lambda^{\prime \prime}_{33k}|^2 & = & -6.0 \pm 2.3.
\end{eqnarray}
Again, the correlations between these constraints are relatively weak, 
so we neglect them henceforth.  
The 1$\sigma$ (2$\sigma$) [3$\sigma$] upper bounds are then:
\begin{eqnarray}
\sum_k |\lambda^{\prime\prime}_{31k}|^2 
& \leq & -1\phantom{.1} \;( \phantom{-}17 )\;[ 35 ] \cr
\sum_k |\lambda^{\prime\prime}_{32k}|^2 
& \leq & -3\phantom{.1} \;( \phantom{-1}7 )\;[ 17 ] \cr
\sum_k |\lambda^{\prime\prime}_{33k}|^2 
& \leq & -3.7 \;( -1.4 )\; [ 0.9 ].
\end{eqnarray}
Recall that the $\lambda^{\prime\prime}$ couplings are antisymmetric 
in the last two indices;  thus, each of the sums above consists of only 
two terms.  
The (2$\sigma$) [3$\sigma$] upper bounds 
on the individual $\lambda^{\prime\prime}$ couplings are then:
\begin{eqnarray}
|\lambda^{\prime\prime}_{321}| & \leq & (2.7)\;[4.1] \cr
|\lambda^{\prime\prime}_{33i}| & \leq & (\phantom{2.7})\;[ 0.96].
\end{eqnarray}
Thus, $\lambda^{\prime\prime}_{331}$ and 
$\lambda^{\prime\prime}_{332}$ are excluded at the 2$\sigma$ level, 
and $\lambda^{\prime\prime}_{321}$ is excluded at the $1 \sigma$
level.  
These bounds significantly improve the 1$\sigma$ bound of 
$|\lambda''_{33k}| < 0.50$ from $R_\ell$
\cite{Allanach:1999ic,Bhattacharyya:1995bw}.

These improvements on the bounds of $\lambda'$ and $\lambda''$ are
a consequence of the fact that while the data prefers a {\it positive}
shift in the right--handed coupling of the $b$, which is non--zero by
2.6$\sigma$, both $\lambda'$ and $\lambda''$ corrections
shift the coupling in the negative direction.  This situation is 
mitigated neither by introducing sfermion mass splittings nor by
increasing $\tan\beta$ \cite{Lebedev:2HDM}.

\section{Bayesian Confidence Intervals for 
$\lambda^{\prime}$ and $\lambda^{\prime \prime}$}
\label{sec:Bayesian}

In the previous section we performed a {\it classical} statistical analysis,
i.e. we performed a fit to the data without any {\it a priori} 
assumptions about the viability 
of the model.  In particular, we made no assumptions about the signs of the 
coupling shifts when fitting the data.  As a consequence, the best-fit
values for the squares of the R-violating couplings were negative, 
resulting in strong $1 \sigma$ and $2 \sigma$ bounds.

An alternate method for calculating confidence levels is to use 
Bayesian statistical analysis.  This technique assumes
that R-violating SUSY is the correct underlying theory, and therefore that
the shifts to the right-handed couplings are only permitted to be negative
and $\dhbLH$ positive.{\footnote{We take the prior probability
for the coupling shifts to be uniform on the region permitted by the 
theory and zero elsewhere.}}
The resulting confidence intervals for the couplings squared are 
positive, and the preferred values are those of the Standard Model (i.e. 
zero).

However, care should be
taken when using these bounds, since they hide the fact that the $\chi^2$ of 
the corresponding fit is quite large even at low confidence levels.  The
probability of these bounds arising as a result of statistical fluctuations
is therefore quite small.

Below we list the $68 \%$  $(95 \%)$ confidence levels from 
the constrained fit:

\begin{eqnarray}
 \dhbR & \geq & -0.0046 \; (-0.010)  \cr
 \dhsR & \geq & -0.031\; (-0.064)\cr
 \dhdR & \geq & -0.061\; (-0.123); 
\end{eqnarray}
The corresponding confidence limits on the couplings are:
\begin{eqnarray}
 |\lambda^{\prime\prime}_{33i}| & \leq & 1.0 \; (1.5)  \cr
 |\lambda^{\prime\prime}_{321}| & \leq & 2.7\; (3.9)
\end{eqnarray}
\begin{eqnarray}
 |\lambda^{\prime}_{i33}| & \leq & 1.4 \; (2.2)  \cr
 |\lambda^{\prime}_{i32}| & \leq & 3.8\; (5.6)\cr
 |\lambda^{\prime}_{i31}| & \leq & 5.2\; (7.6); 
\end{eqnarray}

The best-fit value for $\dhbLH$ is negative.  
However, 
the model generically predicts a positive $\dhbLH$.
The best-fit value of 
$\dhbLH$ consistent with the model is zero;
as a result of this tension, the corresponding $\chi^2$ increases even
further.

To be quantitative concerning the large $\chi^2$ of the constrained fit 
confidence intervals, we present the following example.
The $\chi^2$ corresponding to the $68\%$ and 
$95\%$ confidence intervals for $\dhbR$ are:

\begin{eqnarray}
68\% : & \; \chi^2/{\rm DOF}=26.1/(16-1) \; \rightarrow & \;{\rm probability}=3.7 \% \cr
95\% : & \; \chi^2/{\rm DOF}=33.6/(16-1) \; \rightarrow & \;{\rm probability}=0.4 \% 
\end{eqnarray}

We see explicitly that the bounds obtained using the Bayesian analysis
are weak, but the $\chi^2$ associated with these bounds is uncomfortably 
large.  If the error bars continue to shrink and the central values
stay unchanged, the relevance of the constrained fit bounds must
be questioned.

%%%%%%%%%%%%%%%%%%%%%%%%%%%%%%%%%%%%%%%%%%%%%%%%%%%%%%%%%%%%%%%%%%%%%%%%%%%%%%
\section{Summary and Conclusions}
\label{sec:Conclusion}

We find
that the hadronic $Z$--decay data from LEP and SLD can be used to 
place significant constraints on the size of R--parity violating 
$\lambda^{\prime}$ and $\lambda^{\prime \prime}$ couplings. 
This is possible because the dominant R--violating interactions correct
the couplings of the right--handed down--type quarks only 
while the dominant R--conserving MSSM interactions
correct only the left--handed couplings.
The parity violating asymmetry parameters $A_q$ are particularly 
sensitive to shifts in the right--handed quark couplings while
blind to shifts in the left--handed couplings.
This allows us to constrain the R--violating interactions
independently from the R--conserving sector.

Current data prefer a shift in right--handed
quark couplings {\it opposite} to the direction predicted by the theory.
As a consequence, {\it all} of the R--violating shifts considered in 
this work are excluded at the 1$\sigma$ level.  
In the $\lambda^{\prime}$ case the strongest bound
is on the $\lambda^{\prime}_{i33}$, 
which are excluded at 2$\sigma$ and on which 
we have set the 3$\sigma$ bound
\begin{equation}
|\lambda^{\prime}_{i33}|  \leq 1.4.
\end{equation}
For the $\lambda^{\prime \prime}$ case, 
the $\lambda^{\prime\prime}_{331}$ 
and $\lambda^{\prime\prime}_{332}$
couplings are excluded at the 2$\sigma$ level, and 
$\lambda^{\prime\prime}_{321}$ is excluded at 1$\sigma$.  
The (2$\sigma$) [3$\sigma$] upper bounds are
\begin{eqnarray}
|\lambda^{\prime\prime}_{321}| & \leq & (2.7) \;[ 4.1 ] \cr
|\lambda^{\prime\prime}_{33i}| & \leq & (\phantom{2.7})\;[ 0.96 ].
\end{eqnarray}
All bounds are calculated assuming a common sfermion mass of 100~GeV.
For larger (common) sfermion masses the above bounds may be interpreted 
as bounds on 
$(|\lambda^{\prime}|,|\lambda^{\prime\prime}| ) \times \sqrt{F(x)/F(x_0)}$, 
where $F(x)$ is defined in Eq.~\ref{eq:F} and 
$x_0=\frac{m_t^2}{({\rm 100 GeV})^2}$.  We have also performed a Bayesian
statistical analysis and obtained corresponding confidence levels.

Generically, R--violating interactions {\it reduce} the magnitude of the 
couplings of the the right--handed quarks to the $Z$ and leave the 
left--handed couplings unchanged.  
Current LEP/SLD data prefers shifts which {\it increase} the magnitude of the
right handed coupling, to the extent that even the Standard Model prediction
is in only marginal agreement with the data.  
Future reductions in the experimental uncertainties in the asymmetry 
parameters without changes in the central values
would eventually rule out both the standard model and the MSSM with 
R--violating couplings of either the $\lambda^\prime$ or 
$\lambda^{\prime\prime}$ variety.

%%%%%%%%%%%%%%%%%%%%%%%%%%%%%%%%%%%%%%%%%%%%%%%%%%%%%%%%%%%%%%%%%%%%%%%%%%%%
%\newpage

\acknowledgements

We thank R. Clare and M. Swartz for providing us
with the latest LEPEWWG data including the correlation matrices, and
David Muller for providing us with detailed instructions 
on how to incorporate the SLD measurement of $A_s$ into our fit.
We also gratefully acknowledge
helpful communications with B. Allanach, Y. Nir, P. Rowson, D. Su and
Z. Sullivan.
This work was supported in part (O.L. and W.L.) by the 
U.~S. Department of Energy, grant DE-FG05-92-ER40709, Task A.

%%%%%%%%%%%%%%%%%%%%%%%%%%%%%%%%%%%%%%%%%%%%%%%%%%%%%%%%%%%%%%%%%%%%%%%%%%%%

%%%%%%%%%%%%%%%%%%%%%%%%%%%%%%%%%%%%%%%%%%%%%%%%%%%%%%%%%%%%%%%%%%%%%%%%%%%%%
\onecolumn
\widetext

\appendix

\newcommand{\pole}{\Delta_\epsilon}

\section*{Feynman Integrals}
The integrals we use here are defined explicitly in \cite{Lebedev:1999vc}.
In the approximation $m_Z^2=0$, the one-loop diagrams which appear in this
work are proportional to the following expressions:
\setlength{\unitlength}{1cm}
\begin{eqnarray}
\raisebox{-1cm}{\begin{picture}(2.5,2)(0,0)
\epsfbox[0 0 90 60]{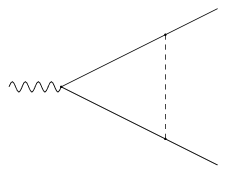}
\end{picture}}
& \propto & 
\left[ (d-2)\,\hat{C}_{24}\left( 0,0,m_Z^2;m_{s},m_{f},m_{f}
                          \right)
     - m_Z^2\,\hat{C}_{23}\left( 0,0,m_Z^2;m_{s},m_{f},m_{f}
                          \right)
\right] 
\cr
& \approx & 
-\frac{1}{(4\pi)^2} 
\left[ \frac{1}{2} \left( \pole - \ln{\frac{m_f^2}{\mu^2}} \right) + f(x)
\right] 
\\
& & \cr
\raisebox{-1cm}{\begin{picture}(2.5,2)(0,0)
\epsfbox[0 0 90 60]{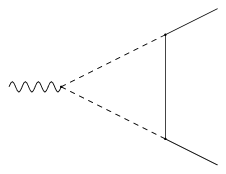}
\end{picture}}
& \propto & 
2\,\hat{C}_{24}\left( 0,0,m_Z^2;m_{f},m_{s},m_{s} \right)
\;\approx\; 
- \frac{1}{(4\pi)^2}
\left[ \frac{1}{2} \left( \pole - \ln{\frac{m_f^2}{\mu^2}} \right) - g(x)
\right] 
\\
& & \cr
\raisebox{-1cm}{\begin{picture}(2.5,2)(0,0)
\epsfbox[0 0 90 60]{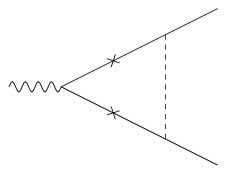}
\end{picture}}
& \propto &  
m_f^2\,\hat{C}_{0}\left( 0,0,m_Z^2;m_{s},m_{f},m_{f} \right) 
\;\approx\; 
-\frac{1}{(4\pi)^2} \left[\, f(x)+ g(x) \,\right]
\\
& & \cr
\raisebox{-1cm}{\begin{picture}(2.5,2)(0,0)
\epsfbox[0 0 90 60]{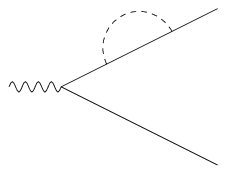}
\end{picture}}
& \propto &  
\hat{B}_{1} \left( 0;m_{f},m_{s} \right)
\;\approx\;
\frac{1}{(4\pi)^2}
\left[ \frac{1}{2} \left( \pole - \ln{\frac{m_f^2}{\mu^2}} \right) - g(x)
\right]
\end{eqnarray}
where 
\begin{eqnarray}
f(x) & = & -\frac{1}{4 (1-x)^2}\left(\,x^2 - 1 - 2 \ln{x} \,\right) \cr
g(x) & = & -\frac{1}{2} \ln{x} + \frac{1}{4 (1-x)^2}
            \left[ -(1-x)(1-3 x) + 2 x^2 \ln{x} \,\right]
\end{eqnarray}
for $x = m_f^2 / m_s^2.$
For $x \longrightarrow 1$ (degenerate scalar and fermion masses),
\begin{eqnarray}
f(x) & \approx & -\frac{1}{2} + \frac{x-1}{6} + \cdots \cr 
g(x) & \approx & - \frac{x-1}{3} + \cdots
\end{eqnarray}
For $x \longrightarrow 0$ (the decoupling limit of heavy scalar masses),
\begin{eqnarray}
f(x) & \approx & \phantom{-}\frac{1}{2} \ln{x} + \frac{1}{4} + \cdots \cr 
g(x) & \approx & - \frac{1}{2} \ln{x}  -  \frac{1}{4} + \cdots 
     \;\sim\; -f(x).
\end{eqnarray}

%%%%%%%%%%%%%%%%%%%%%%%%%%%%%%%%%%%%%%%%%%%%%%%%%%%%%%%%%%%%%%%%%%%%%%%%%%%%
\widetext

\begin{table}[p]
\begin{center}
\begin{tabular}{|c|c|c|c|}
Observable & Reference & Measured Value & ZFITTER Prediction \\
\hline\hline
\multicolumn{2}{|l|}{\underline{$Z$ lineshape variables}} & & \\
$m_Z$                & \cite{MNICH:99} & $91.1872 \pm 0.0021$ GeV & input \\
$\Gamma_Z$           & \cite{MNICH:99} & $2.4944 \pm 0.0024$ GeV  & --- \\
$\sigma_{\rm had}^0$ & \cite{MNICH:99} & $41.544 \pm 0.037$ nb    & --- \\
$R_e$                & \cite{MNICH:99} & $20.803 \pm 0.049$       & --- \\
$R_\mu$              & \cite{MNICH:99} & $20.786 \pm 0.033$       & --- \\
$R_\tau$             & \cite{MNICH:99} & $20.764 \pm 0.045$       & --- \\
$A_{\rm FB}(e   )$   & \cite{MNICH:99} & $0.0145 \pm 0.0024$      & $0.0152$ \\
$A_{\rm FB}(\mu )$   & \cite{MNICH:99} & $0.0167 \pm 0.0013$      & --- \\
$A_{\rm FB}(\tau)$   & \cite{MNICH:99} & $0.0188 \pm 0.0017$      & --- \\
$R_{\nu/e}$          &                 & $1.9755 \pm 0.0080$      & $1.9916$ \\
\hline
\multicolumn{2}{|l|}{\underline{$\tau$ polarization at LEP}} & & \\
$A_e$        & \cite{MNICH:99} & $0.1483 \pm 0.0051$   & $0.1423$ \\ 
$A_\tau$     & \cite{MNICH:99} & $0.1424 \pm 0.0044$   & ---  \\
\hline
\multicolumn{2}{|l|}{\underline{SLD left--right asymmetries}} & & \\
$A_{LR}$     & \cite{SLD:99} & $0.15108 \pm 0.00218$ & $0.1423$ \\
$A_e$        & \cite{SLD:99} & $0.1558  \pm 0.0064$  & $0.1423$ \\
$A_{\mu}$    & \cite{SLD:99} & $0.137   \pm 0.016$   & ---  \\
$A_{\tau}$   & \cite{SLD:99} & $0.142   \pm 0.016$   & ---  \\
\hline
\multicolumn{2}{|l|}{\underline{light quark flavor}} & & \\
$R_s^{\prime *}$  [OPAL]   & \cite{OPAL:97} & $0.392  \pm 0.062$  & $0.360$\\
$A_{\rm FB}^*(s)$ [OPAL]   & \cite{OPAL:97} & $0.075  \pm 0.029$  & $0.100$\\
$A_{\rm FB}^*(u)$ [OPAL]   & \cite{OPAL:97} & $0.086  \pm 0.037$  & $0.071$\\
$A_{\rm FB}^{**}(s)$ [DELPHI]
                           & \cite{DELPHI:99} & $0.1008\pm 0.0120$ & $0.1006$\\
$A_s^{*}$ [SLD]            & \cite{As-SLD:99} & $0.85  \pm 0.092$  & $0.935$ \\
\hline
\multicolumn{2}{|l|}{\underline{heavy quark flavor}} & & \\
$R_b$           & \cite{MNICH:99} & $0.21642 \pm 0.00073$ & $0.21583$ \\
$R_c$           & \cite{MNICH:99} & $0.1674  \pm 0.0038$  & $0.1722$ \\
$A_{\rm FB}(b)$ & \cite{MNICH:99} & $0.0988  \pm 0.0020$  & $0.0997$ \\
$A_{\rm FB}(c)$ & \cite{MNICH:99} & $0.0692  \pm 0.0037$  & $0.0711$ \\
$A_b$           & \cite{MNICH:99} & $0.911   \pm 0.025$   & $0.934$ \\
$A_c$           & \cite{MNICH:99} & $0.630   \pm 0.026$   & $0.666$ \\
\end{tabular}
\caption{LEP/SLD observables and their Standard Model predictions.
The ratio 
$R_{\nu/e} = \Gamma_{\nu\bar{\nu}}/\Gamma_{e^+e^-}$
was calculated from the $Z$--lineshape observables.
The Standard Model predictions were calculated using ZFITTER v.6.21 
\protect\cite{ZFITTER:99} with $m_t = 174.3$~GeV \protect\cite{TOPMASS:99},
$m_H = 300$~GeV, and $\alpha_s(m_Z) = 0.120$ as input.}
\label{LEP-SLD-DATA}
\end{center}
\end{table}

\medskip

\widetext

\begin{table}[ht]
\begin{center}
\begin{tabular}{|c|ccccccccc|}
& $m_Z$     & $\Gamma_Z$     & $\sigma_{\rm had}^0$
& $R_e$     & $R_\mu$     & $R_\tau$ 
& $A_{\rm FB}(e)$ & $A_{\rm FB}(\mu)$ & $A_{\rm FB}(\tau)$ \\
\hline
$m_Z$ 
& $1.000$            & $-0.008$           & $-0.050$ 
& $\phantom{-}0.073$ & $\phantom{-}0.001$ & $\phantom{-}0.002$
& $-0.015$           & $\phantom{-}0.046$ & $\phantom{-}0.034$ \\
$\Gamma_Z$
&                    & $\phantom{-}1.000$ & $-0.284$ 
& $-0.006$           & $\phantom{-}0.008$ & $\phantom{-}0.000$ 
& $-0.002$           & $\phantom{-}0.002$ & $-0.003$ \\
$\sigma_{\rm had}^0$
&                    &                    & $\phantom{-}1.000$
& $\phantom{-}0.109$ & $\phantom{-}0.137$ & $\phantom{-}0.100$ 
& $\phantom{-}0.008$ & $\phantom{-}0.001$ & $\phantom{-}0.007$ \\
$R_e$
&                    &                    &
& $\phantom{-}1.000$ & $\phantom{-}0.070$ & $\phantom{-}0.044$ 
& $-0.356$           & $\phantom{-}0.023$ & $\phantom{-}0.016$ \\
$R_\mu$
&                    &                    &
&                    & $\phantom{-}1.000$ & $\phantom{-}0.072$ 
& $\phantom{-}0.005$ & $\phantom{-}0.006$ & $\phantom{-}0.004$ \\
$R_\tau$
&                    &                    &
&                    &                    & $\phantom{-}1.000$
& $\phantom{-}0.003$ & $-0.003$           & $\phantom{-}0.010$ \\
$A_{\rm FB}(e)$
&                    &                    & 
&                    &                    &
& $\phantom{-}1.000$ & $-0.026$           & $-0.020$ \\
$A_{\rm FB}(\mu)$ 
&                    &                    & 
&                    &                    & 
&                    & $\phantom{-}1.000$ & $\phantom{-}0.045$ \\
$A_{\rm FB}(\tau)$
&                    &                    & 
&                    &                    & 
&                    &                    & $\phantom{-}1.000$ \\
\end{tabular}
\caption{The correlation of the $Z$ lineshape variables at LEP.}
\label{lineshape-correlations}
\end{center}
\end{table}

\begin{table}[ht]
\begin{center}
\begin{tabular}{|c|cccccc|}
& $R_b$              & $R_c$     
& $A_{\rm FB}(b)$    & $A_{\rm FB}(c)$     
& $A_b$              & $A_c$ \\
\hline
$R_b$ 
& $1.00$            & $-0.14$           & $-0.03$ 
& $\phantom{-}0.01$ & $-0.03$           & $\phantom{-}0.02$ \\
$R_c$
&                   & $\phantom{-}1.00$ & $\phantom{-}0.05$ 
& $-0.05$           & $\phantom{-}0.02$ & $-0.02$ \\
$A_{\rm FB}(b)$
&                   &                   & $\phantom{-}1.00$
& $\phantom{-}0.09$ & $\phantom{-}0.02$ & $\phantom{-}0.00$ \\
$A_{\rm FB}(c)$
&                   &                   &
& $\phantom{-}1.00$ & $-0.01$           & $\phantom{-}0.03$ \\
$A_b$
&                   &                   &
&                   & $\phantom{-}1.00$ & $\phantom{-}0.15$ \\ 
$A_c$
&                   &                   &
&                   &                   & $\phantom{-}1.00$ \\
\end{tabular}
\caption{The correlation of the heavy flavor variables from LEP/SLD.}
\label{heavy-correlations}
\end{center}
\end{table}

\begin{table}[ht]
\begin{center}
\begin{tabular}{|c|rrrrr|}
& $\dss$ & $\dhdR$ & $\dhsR$ & $\dhbR$ & $\dhbLH$ \\
\hline\hline
$\dss$   & $1.00$ & $0.01$  & $-0.06$ & $-0.42$ & $-0.15$ \\
$\dhdR$  &        &  $1.00$ & $-0.30$ &  $0.09$ & $-0.75$ \\
$\dhsR$  &        &         &  $1.00$ &  $0.05$ & $-0.22$ \\
$\dhbR$  &        &         &         &  $1.00$ &  $0.30$ \\
$\dhbLH$ &        &         &         &         &  $1.00$ 
\end{tabular}
\caption{The correlation matrix of the fit parameters.}
\label{fit-correlation}
\end{center}
\end{table}

%%%%%%%%%%%%%%%%%%%%%%%%%%%%%%%%%%%%%%%%%%%%%%%%%%%%%%%%%%%%%%%%%%%%%%%%%%%%%

%%%%%%%%%%%%%%%%%%%%%%%%%%%%%%%%%%%%%%%%%%%%%%%%%%%%%%%%%%%%%%%%%%%%%%%%%%%%%
\begin{figure}[ht]
\begin{center}
\unitlength=1cm
\begin{picture}(17,11)(0,0)
\unitlength=1mm
%\zahyou{17}{11}
\put(13,77){\vector(1,0){8}}
\put(15,71){$Q$}
\put(40,92){\vector(2,1){6}}
\put(40,94){$p$}
\put(40,68){\vector(2,-1){6}}
\put(40,64){$q$}
\put(30,55){(a)}
\put(85,55){(b)}
\put(140,55){(c)}
\put(60,5){(d)}
\put(115,5){(e)}
\put(8,80){$Z$}
\put(42,80){$t_L$}
\put(63,80){$Z$}
\put(97,80){$\tilde{e}_{i_L}$}
\put(118,80){$Z$}
\put(152,80){$\tilde{e}_{i_L}$}
\put(35,31){$Z$}
\put(90,31){$Z$}
\put(50,94){$d_{k_R}$}
\put(50,66){$d_{k_R}$}
\put(106,94){$d_{k_R}$}
\put(106,66){$d_{k_R}$}
\put(160,94){$d_{k_R}$}
\put(160,66){$d_{k_R}$}
\put(78,45){$d_{k_R}$}
\put(78,16){$d_{k_R}$}
\put(133,45){$d_{k_R}$}
\put(133,16){$d_{k_R}$}
\put(26,88){$\tilde{e}_{i_L}$}
\put(26,72){$\tilde{e}_{i_L}$}
\put(81,88){$t_L$}
\put(81,72){$t_L$}
\put(132,86){$t_R$}
\put(132,74){$t_R$}
\put(142,90){$t_L$}
\put(142,70){$t_L$}
\put(67,30){$\tilde{e}_{i_L}$}
\put(48,37){$d_{k_R}$}
\put(60,41){$t_L$}
\put(103,26){$d_{k_R}$}
\put(121,32){$\tilde{e}_{i_L}$}
\put(115,20){$t_L$}
\epsfbox[0 470 480 780]{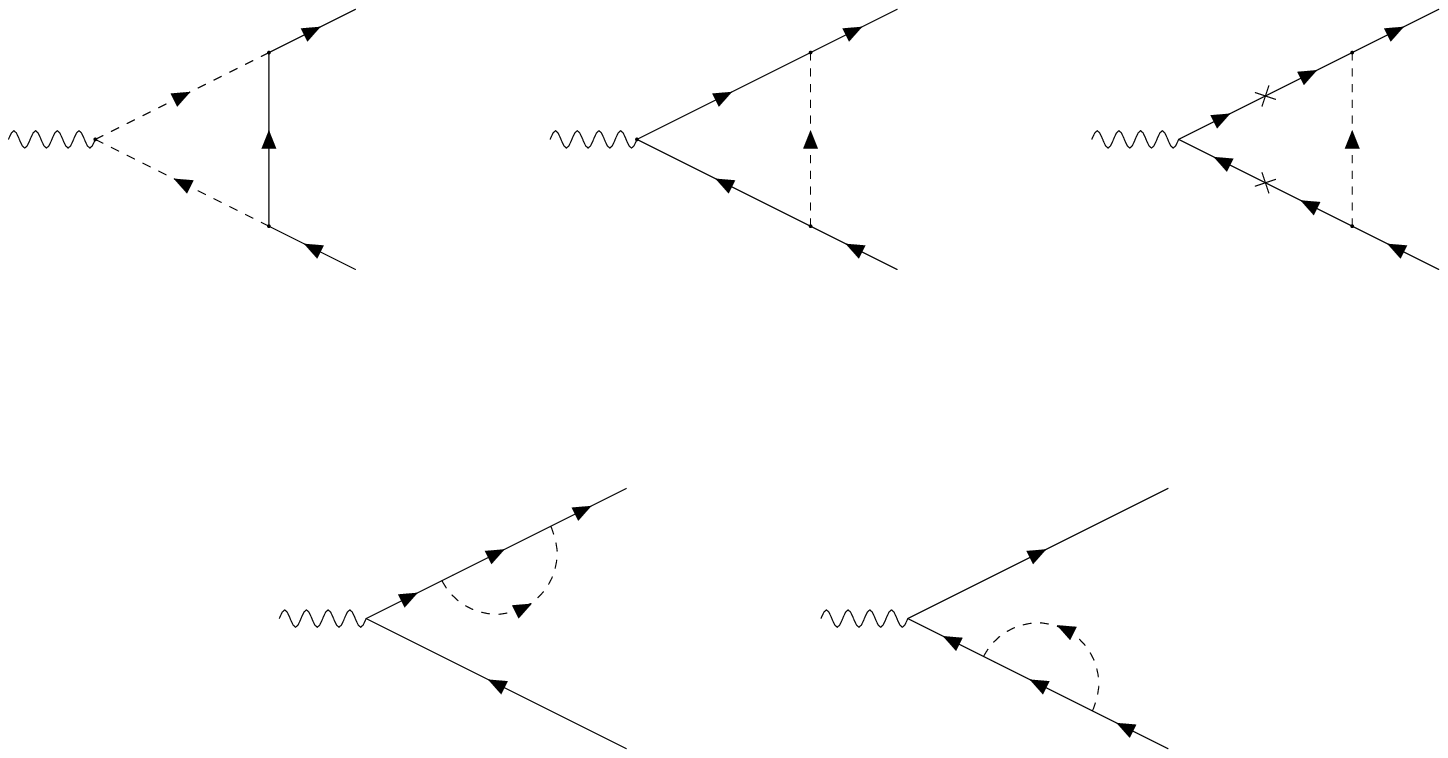}
\end{picture}
\caption{One--loop corrections to
$Z\rightarrow d_{k_R}\bar{d}_{k_R}$ involving the 
R--parity violating $\lambda^{\prime}$ couplings.} 
\label{Zdecaylambdaprime}
\end{center}
\end{figure}

%%%%%%%%%%%%%%%%%%%%%%%%%%%%%%%%%%%%%%%%%%%%%%%%%%%%%%%%%%%%%%%%%%%%%%%%%%%%
\newpage

\begin{figure}[ht]
\begin{center}
\unitlength=1cm
\begin{picture}(17,11)(0,0)
\unitlength=1mm
%\zahyou{17}{11}
\put(13,77){\vector(1,0){8}}
\put(15,71){$Q$}
\put(40,92){\vector(2,1){6}}
\put(40,94){$p$}
\put(40,68){\vector(2,-1){6}}
\put(40,64){$q$}
\put(30,55){(a)}
\put(85,55){(b)}
\put(140,55){(c)}
\put(60,5){(d)}
\put(115,5){(e)}
\put(8,80){$Z$}
\put(42,80){$t_R$}
\put(63,80){$Z$}
\put(97,80){$\tilde{d}_{k_R}$}
\put(118,80){$Z$}
\put(152,80){$\tilde{d}_{k_R}$}
\put(35,31){$Z$}
\put(90,31){$Z$}
\put(50,94){$d_{j_R}$}
\put(50,66){$d_{j_R}$}
\put(106,94){$d_{j_R}$}
\put(106,66){$d_{j_R}$}
\put(160,94){$d_{j_R}$}
\put(160,66){$d_{j_R}$}
\put(78,45){$d_{j_R}$}
\put(78,16){$d_{j_R}$}
\put(133,45){$d_{j_R}$}
\put(133,16){$d_{j_R}$}
\put(26,88){$\tilde{d}_{k_R}$}
\put(26,72){$\tilde{d}_{k_R}$}
\put(81,88){$t_R$}
\put(81,72){$t_R$}
\put(132,86){$t_L$}
\put(132,74){$t_L$}
\put(142,90){$t_R$}
\put(142,70){$t_R$}
\put(54,46){$\tilde{d}_{k_R}$}
\put(48,37){$d_{j_R}$}
\put(63,34){$t_R$}
\put(103,26){$d_{j_R}$}
\put(111,15){$\tilde{d}_{k_R}$}
\put(119,26){$t_R$}
\epsfbox[0 470 480 780]{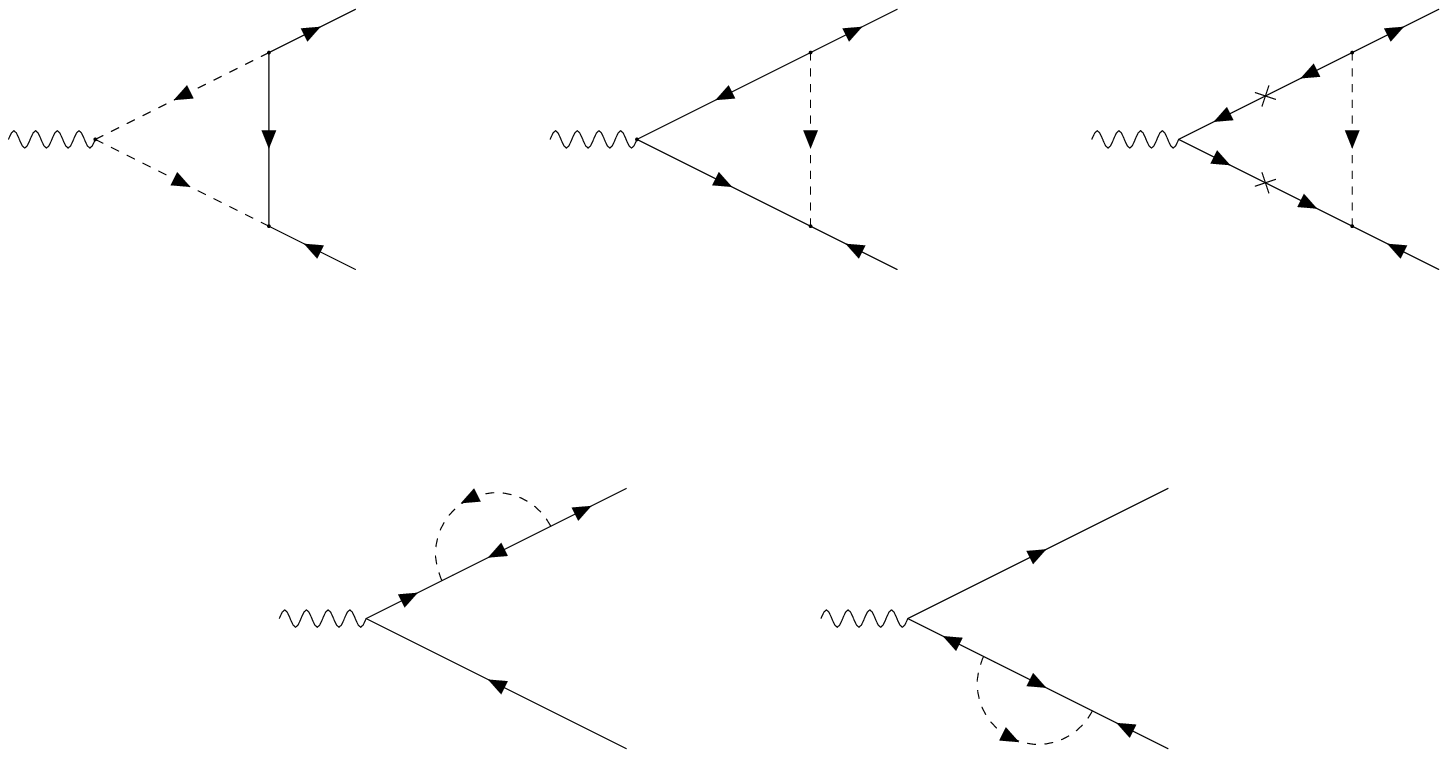}
\end{picture}
\caption{One--loop corrections to
$Z\rightarrow d_{j_R}\bar{d}_{j_R}$ involving the 
R--parity violating $\lambda^{\prime \prime}$ couplings.} 
\label{Zdecaylambdadoubleprime}
\end{center}
\end{figure}

%%%%%%%%%%%%%%%%%%%%%%%%%%%%%%%%%%%%%%%%%%%%%%%%%%%%%%%%%%%%%%%%%%%%%%%%%%%%
\newpage

\begin{figure}[ht]
\begin{center}
\unitlength=1cm
\begin{picture}(13,10)(0,0)
\unitlength=1mm
%\zahyou{13}{10}
\put(40,55){(a)}
\put(96,55){(b)}
\put(40,5){(c)}
\put(96,5){(d)}
\put(17,78){$Z$}
\put(72,78){$Z$}
\put(17,29){$Z$}
\put(72,29){$Z$}
\put(50,78){$\tilde{u}_L$}
\put(46,27){$\tilde{u}_L$}
\put(102,31){$\tilde{u}_L$}
\put(105,78){$\tilde{\chi}^-$}
\put(40,39){$\tilde{\chi}^-$}
\put(96,19){$\tilde{\chi}^-$}
\put(30,35){$d_L$}
\put(86,24){$d_L$}
\put(59,92){$d_L$}
\put(114,92){$d_L$}
\put(59,64){$d_L$}
\put(114,64){$d_L$}
\put(59,43){$d_L$}
\put(114,43){$d_L$}
\put(59,15){$d_L$}
\put(114,15){$d_L$}
\put(36,87){$\tilde{\chi}^-$}
\put(36,69){$\tilde{\chi}^-$}
\put(92,87){$\tilde{u}_L$}
\put(92,69){$\tilde{u}_L$}
\epsfbox[0 0 360 290]{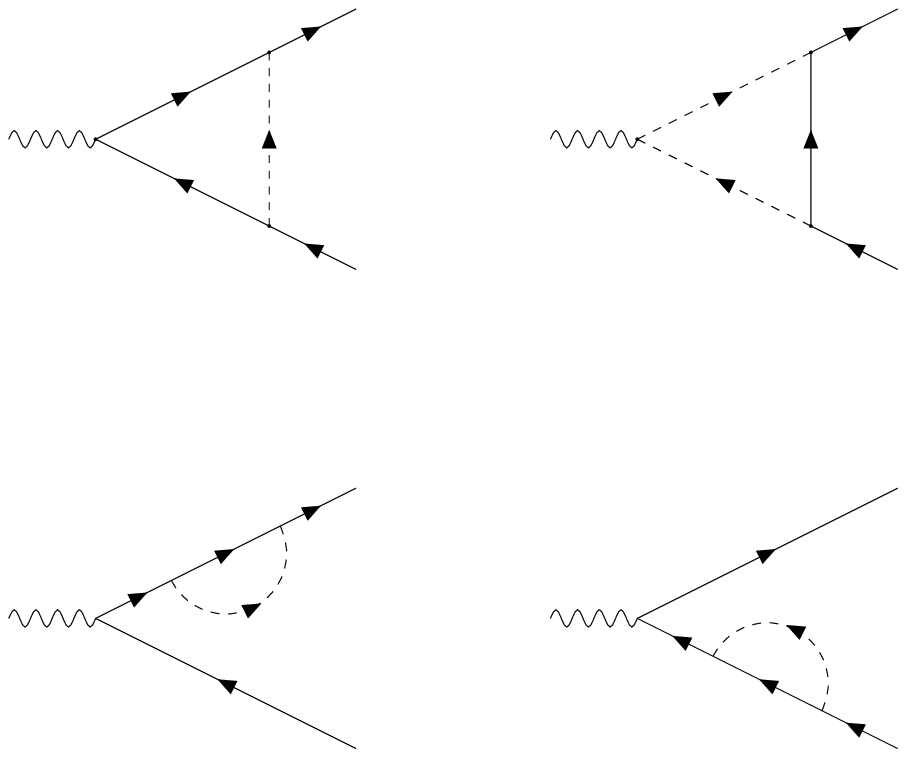}
\end{picture}
\caption{Examples of leading 1PI R--parity conserving
chargino--sfermion contributions subsumed into $\dss$.
There are analogous diagrams with $u_L$ quark final states and
with lepton final states.}
\label{CharginoSfermion}
\end{center}
\end{figure}

%%%%%%%%%%%%%%%%%%%%%%%%%%%%%%%%%%%%%%%%%%%%%%%%%%%%%%%%%%%%%%%%%%%%%%%%%%%%

\begin{figure}[ht]
\begin{center}
\unitlength=1cm
\begin{picture}(13,15)(0,0)
\unitlength=1mm
%\zahyou{13}{15}
\put(40,105){(a)}
\put(96,105){(b)}
\put(40,55){(c)}
\put(96,55){(d)}
\put(68,5){(e)}
\put(16,80){$Z$}
\put(71,80){$Z$}
\put(16,129){$Z$}
\put(71,129){$Z$}
\put(44,31){$Z$}
\put(86,16){$b_L$}
\put(86,46){$b_L$}
\put(59,94){$b_L$}
\put(114,94){$b_L$}
\put(59,66){$b_L$}
\put(114,66){$b_L$}
\put(59,143){$b_L$}
\put(114,143){$b_L$}
\put(59,115){$b_L$}
\put(114,115){$b_L$}
\put(36,90){$\tilde{\chi}^-$}
\put(36,71){$\tilde{\chi}^-$}
\put(50,80){$\tilde{t}_R$}
\put(92,90){$t_R$}
\put(92,71){$t_R$}
\put(105,80){$H^-$}
\put(36,138){$H^-$}
\put(36,119){$H^-$}
\put(50,128){$t_R$}
\put(92,138){$\tilde{t}_R$}
\put(92,120){$\tilde{t}_R$}
\put(105,128){$\tilde{\chi}^-$}
\put(58,26){$t_L$}
\put(58,39){$t_L$}
\put(68,22){$t_R$}
\put(68,43){$t_R$}
\put(78,31){$H^-$}
\epsfbox[0 330 360 780]{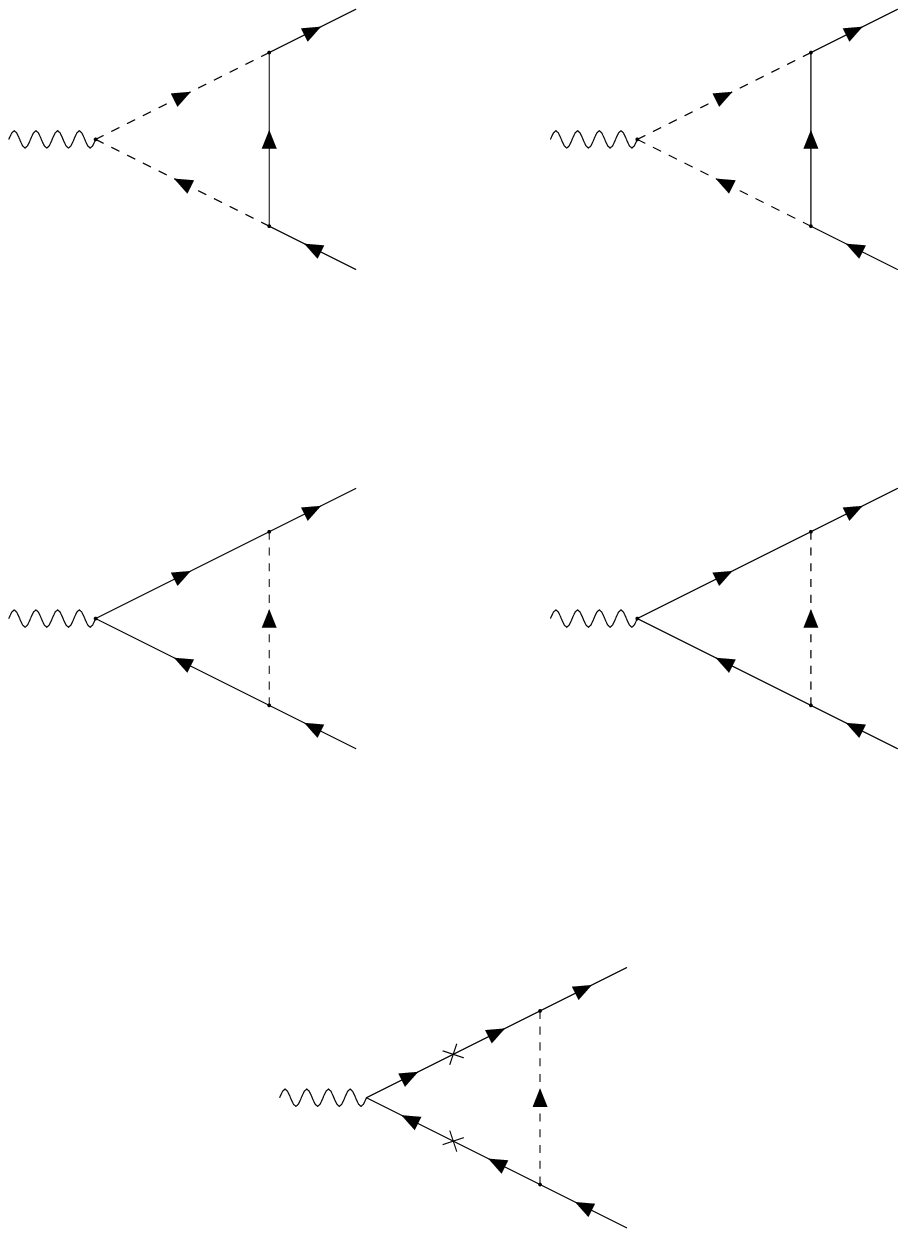}
\end{picture}
\caption{Leading 1PI R--parity conserving
contributions specific to $\dhbLH$.}  
\label{TopHiggsLoop}
\end{center}
\end{figure}

%%%%%%%%%%%%%%%%%%%%%%%%%%%%%%%%%%%%%%%%%%%%%%%%%%%%%%%%%%%%%%%%%%%%%%%%%%%%

\begin{figure}[ht]
\begin{center}
\unitlength=1cm
\begin{picture}(17,6)(0,0)
\unitlength=1mm
%\zahyou{17}{6}
\put(30,5){(a)}
\put(85,5){(b)}
\put(140,5){(c)}
\put(8,33){$Z$}
\put(63,33){$Z$}
\put(118,33){$Z$}
\put(42,33){$\tilde{g}$}
\put(50,48){$q$}
\put(50,19){$q$}
\put(105,48){$q$}
\put(105,19){$q$}
\put(159,48){$q$}
\put(159,19){$q$}
\put(29,42){$\tilde{q}$}
\put(29,25){$\tilde{q}$}
\put(78,39){$q$}
\put(87,44){$\tilde{g}$}
\put(94,32){$\tilde{q}$}
\put(133,28){$q$}
\put(142,23){$\tilde{g}$}
\put(148,36){$\tilde{q}$}
\epsfbox[0 600 480 780]{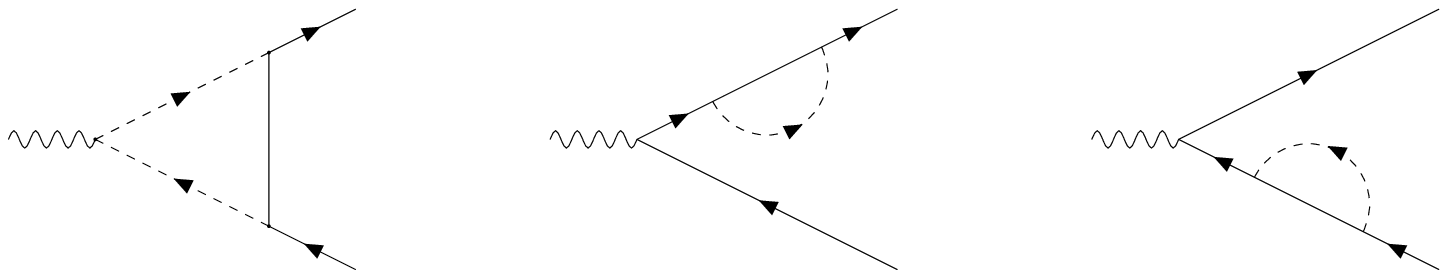}
\end{picture}
\caption{Gluino--squark corrections to the $Zq\bar{q}$ vertex.} 
\label{GluinoSquark}
\end{center}
\end{figure}

%%%%%%%%%%%%%%%%%%%%%%%%%%%%%%%%%%%%%%%%%%%%%%%%%%%%%%%%%%%%%%%%%%%%%%%%%%%%

\begin{figure}[ht]
\begin{center}
\unitlength=1cm
\begin{picture}(17,6)(0,0)
\unitlength=1mm
%\zahyou{17}{6}
\put(30,5){(a)}
\put(85,5){(b)}
\put(140,5){(c)}
\put(8,33){$Z$}
\put(63,33){$Z$}
\put(118,33){$Z$}
\put(42,33){$\tilde{\chi}^0$}
\put(50,48){$q$}
\put(50,19){$q$}
\put(105,48){$q$}
\put(105,19){$q$}
\put(159,48){$q$}
\put(159,19){$q$}
\put(29,42){$\tilde{q}$}
\put(29,25){$\tilde{q}$}
\put(78,39){$q$}
\put(87,44){$\tilde{\chi}^0$}
\put(94,32){$\tilde{q}$}
\put(133,28){$q$}
\put(142,23){$\tilde{\chi}^0$}
\put(148,36){$\tilde{q}$}
\epsfbox[0 600 480 780]{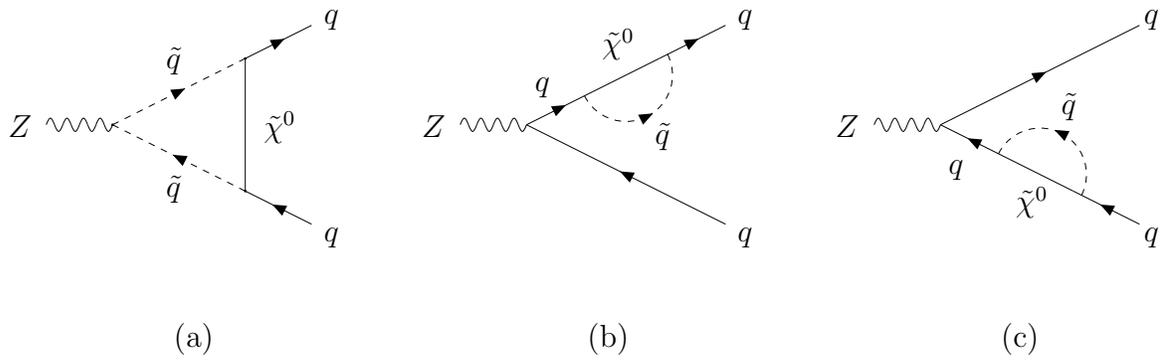}
\end{picture}
\caption{Neutralino--squark corrections to the $Zq\bar{q}$ vertex.} 
\label{Neutralino}
\end{center}
\end{figure}

%%%%%%%%%%%%%%%%%%%%%%%%%%%%%%%%%%%%%%%%%%%%%%%%%%%%%%%%%%%%%%%%%%%%%%%%%%%%

%%%%%%%%%%%%%%%%%%%%%%%%%%%%%%%%%%%%

\begin{figure}[p]
\centering
\unitlength=1cm
\begin{picture}(12,10)
\unitlength=1mm
%\zahyou{12}{10}
\put(60,5){$\dss$}
\put(1,53){$\dhdR$}
\put(33,72){$R_s^{\prime *}$}
\put(35,84){$A_e({\rm SLD})$}
\put(46,77){$A_{\rm LR}$}
\put(40,37){$A_e({\rm LEP})$}
\put(69,29){$A_{\rm FB}(c)$}
\put(71,22){$A_{\rm FB}(e)$}
\put(86,29){$A_{\rm FB}^*(s)$}
\put(87,80){$A_{\rm FB}^{**}(s)$}
\put(84,41){$A_{\rm FB}^*(u)$}
\epsfbox[0 0 300 270]{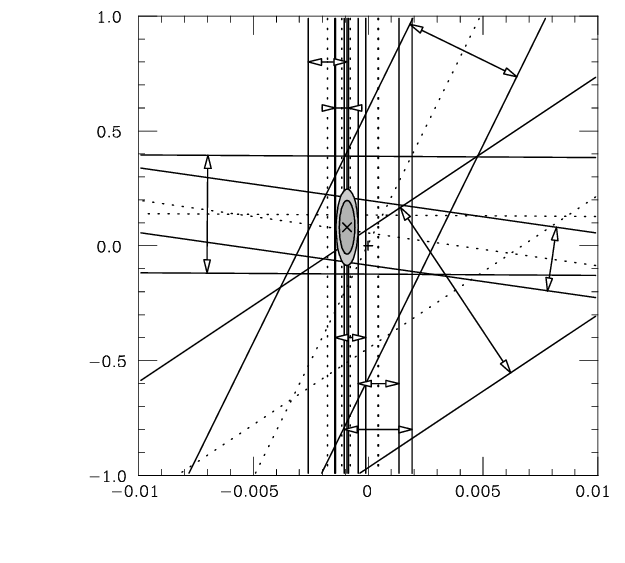}
\end{picture}
\caption{Constraints in the $\dss$--$\dhdR$ plane from various observables.}
\label{FIG12}
\end{figure}

%%%%%%%%%%%%%%%%%%%%%%%%%%%%%%%%%%%%

\begin{figure}[p]
\centering
\unitlength=1cm
\begin{picture}(12,10)
\unitlength=1mm
%\zahyou{12}{10}
\put(60,5){$\dss$}
\put(1,53){$\dhsR$}
\put(35,84){$A_e({\rm SLD})$}
\put(46,77){$A_{\rm LR}$}
\put(40,37){$A_e({\rm LEP})$}
\put(69,29){$A_{\rm FB}(c)$}
\put(71,22){$A_{\rm FB}(e)$}
\put(79,64){$A_{\rm FB}^*(s)$}
\put(81,40){$A_{\rm FB}^{**}(s)$}
\put(31,49){$A_s^*$}
\epsfbox[0 0 300 270]{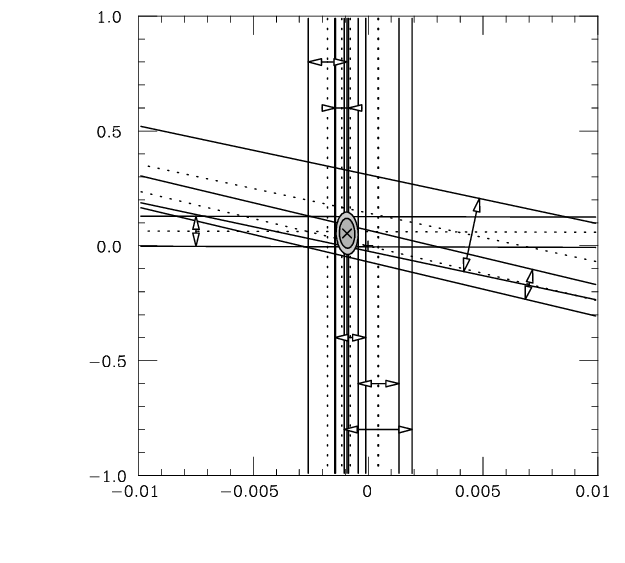}
\end{picture}
\caption{Constraints in the $\dss$--$\dhsR$ plane from various observables.}
\label{FIG13}
\end{figure}

%%%%%%%%%%%%%%%%%%%%%%%%%%%%%%%%%%%%

\begin{figure}[p]
\centering
\unitlength=1cm
\begin{picture}(12,10)
\unitlength=1mm
%\zahyou{12}{10}
\put(60,5){$\dss$}
\put(1,53){$\dhbR$}
\put(71,84){$A_{\rm FB}(e)$}
\put(69,76){$A_{\rm FB}(c)$}
\put(40,38){$A_e({\rm LEP})$}
\put(46,30){$A_{\rm LR}$}
\put(35,22){$A_e({\rm SLD})$}
\put(88,66){$A_b$}
\put(33,50){$R_b$}
\put(77.5,24){$A_{\rm FB}(b)$}
\epsfbox[0 0 300 270]{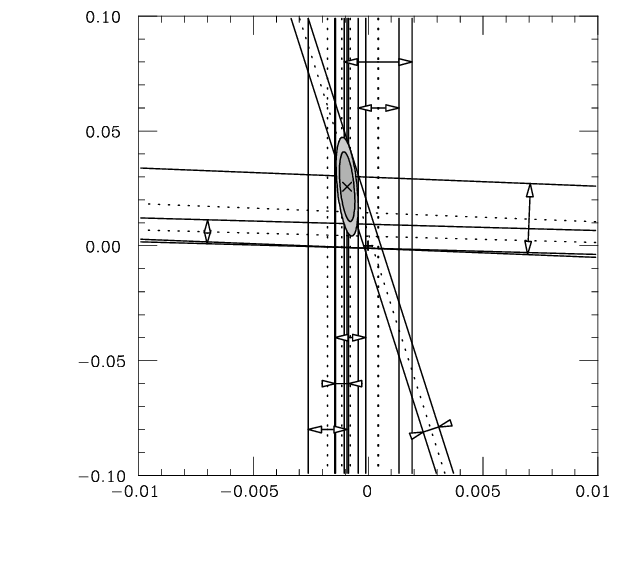}
\end{picture}
\caption{Constraints in the $\dss$--$\dhbR$ plane from various observables.}
\label{FIG14}
\end{figure}

%%%%%%%%%%%%%%%%%%%%%%%%%%%%%%%%%%%%

\begin{figure}[p]
\centering
\unitlength=1cm
\begin{picture}(12,10)
\unitlength=1mm
%\zahyou{12}{10}
\put(60,5){$\dss$}
\put(1,53){$\dhbLH$}
\put(35,84){$A_e({\rm SLD})$}
\put(46,77){$A_{\rm LR}$}
\put(40,68){$A_e({\rm LEP})$}
\put(68,61){$A_{\rm FB}(c)$}
\put(70.5,29){$A_{\rm FB}(e)$}
\put(69,22){$A_{\rm FB}(b)$}
\put(88,85){$A_b$}
\put(90,58){$R_b$}
\put(33,29){$R_c$}
\epsfbox[0 0 300 270]{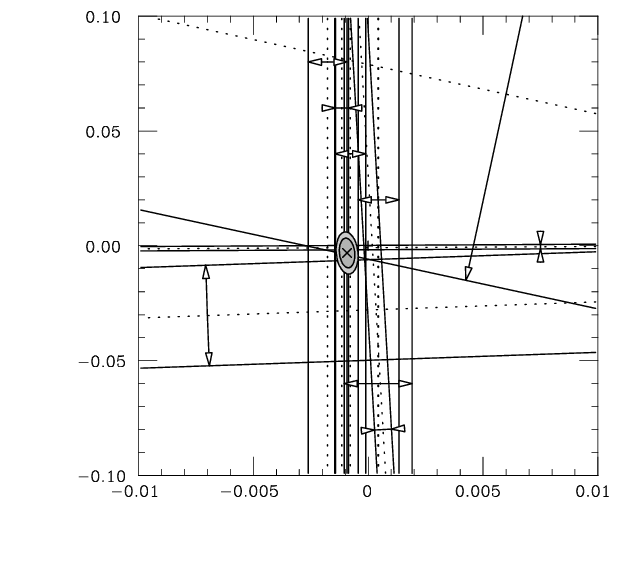}
\end{picture}
\caption{Constraints in the $\dss$--$\dhbLH$ plane from various observables.}
\label{FIG15}
\end{figure}

%%%%%%%%%%%%%%%%%%%%%%%%%%%%%%%%%%%%

\begin{figure}[p]
\centering
\unitlength=1cm
\begin{picture}(12,10)
\unitlength=1mm
%\zahyou{12}{10}
\put(59,5){$\dhdR$}
\put(1,53){$\dhsR$}
\put(78,32){$R_s^{\prime *}$}
\put(80,77){$A_{\rm FB}^*(s)$}
\put(44,80){$A_{\rm FB}^*(u)$}
\put(29,62){$A_s^*$}
\put(34,43){$A_{\rm FB}^{**}(s)$}
\epsfbox[0 0 300 270]{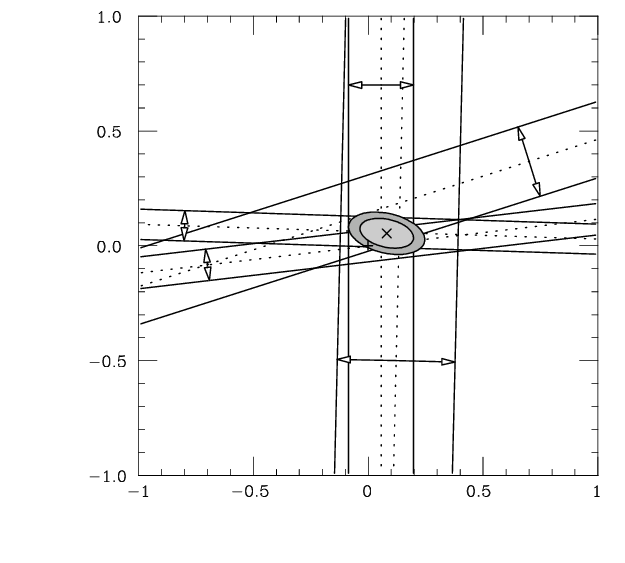}
\end{picture}
\caption{Constraints in the $\dhdR$--$\dhsR$ plane from various observables.}
\label{FIG23}
\end{figure}

%%%%%%%%%%%%%%%%%%%%%%%%%%%%%%%%%%%%

\begin{figure}[p]
\centering
\unitlength=1cm
\begin{picture}(12,10)
\unitlength=1mm
%\zahyou{12}{10}
\put(59,5){$\dhdR$}
\put(1,53){$\dhbR$}
\put(44,80){$A_{\rm FB}^*(u)$}
\put(78,35){$R^{\prime *}_s$}
\put(78,24){$R_c$}
\put(42.5,25){$R_b$}
\put(36,46){$A_{\rm FB}(b)$}
\put(88,67){$A_b$}
\epsfbox[0 0 300 270]{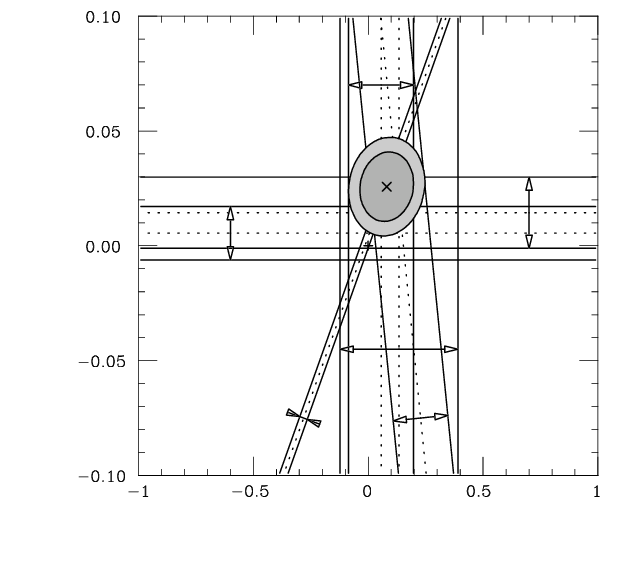}
\end{picture}
\caption{Constraints in the $\dhdR$--$\dhbR$ plane from various observables.}
\label{FIG24}
\end{figure}

%%%%%%%%%%%%%%%%%%%%%%%%%%%%%%%%%%%%
\begin{figure}[p]
\centering
\unitlength=1cm
\begin{picture}(12,10)
\unitlength=1mm
%\zahyou{12}{10}
\put(59,5){$\dhdR$}
\put(1,53){$\dhbLH$}
\put(44,77){$A_{\rm FB}^*(u)$}
\put(79,21){$R^{\prime *}_s$}
\put(81,64){$R_c$}
\put(39,57){$R_b$}
\epsfbox[0 0 300 270]{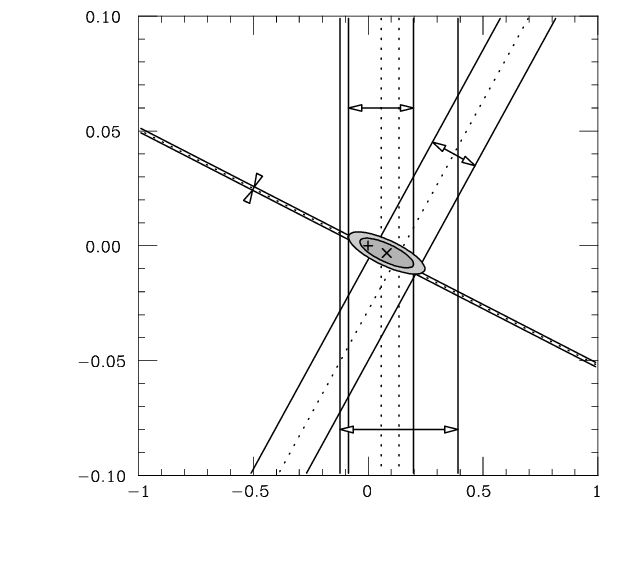}
\end{picture}
\caption{Constraints in the $\dhdR$--$\dhbR$ plane from various observables.}
\label{FIG25}
\end{figure}

%%%%%%%%%%%%%%%%%%%%%%%%%%%%%%%%%%%%

\begin{figure}[p]
\centering
\unitlength=1cm
\begin{picture}(12,10)
\unitlength=1mm
%\zahyou{12}{10}
\put(59,5){$\dhsR$}
\put(1,53){$\dhbR$}
\put(36,46){$A_{\rm FB}(b)$}
\put(88,67){$A_b$}
\put(75,76){$A_{\rm FB}^*(s)$}
\put(46,84){$A_{\rm FB}^{**}(s)$}
\put(68,33){$A^*_s$}
\put(77,24){$R_c$}
\put(43,26){$R_b$}
\epsfbox[0 0 300 270]{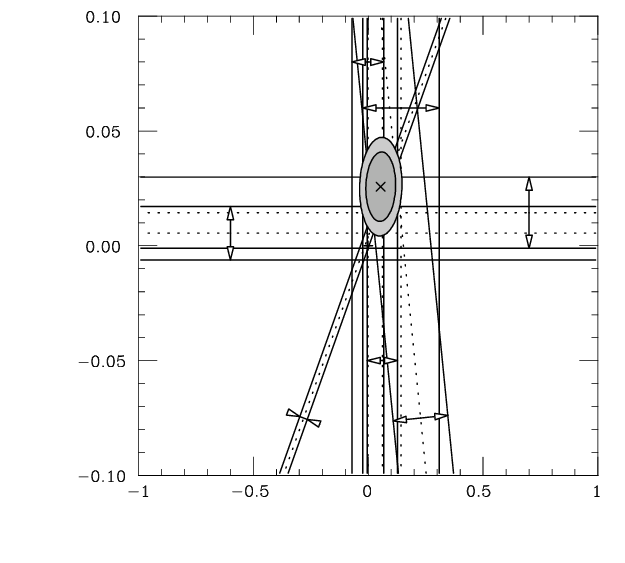}
\end{picture}
\caption{Constraints in the $\dhdR$--$\dhbR$ plane from various observables.}
\label{FIG34}
\end{figure}

%%%%%%%%%%%%%%%%%%%%%%%%%%%%%%%%%%%%

\begin{figure}[p]
\centering
\unitlength=1cm
\begin{picture}(12,10)
\unitlength=1mm
%\zahyou{12}{10}
\put(59,5){$\dhsR$}
\put(1,53){$\dhbLH$}
\put(75,22){$A_{\rm FB}^*(s)$}
\put(46,76){$A_{\rm FB}^{**}(s)$}
\put(68,30){$A^*_s$}
\put(43,27){$R_c$}
\put(39,57){$R_b$}
\epsfbox[0 0 300 270]{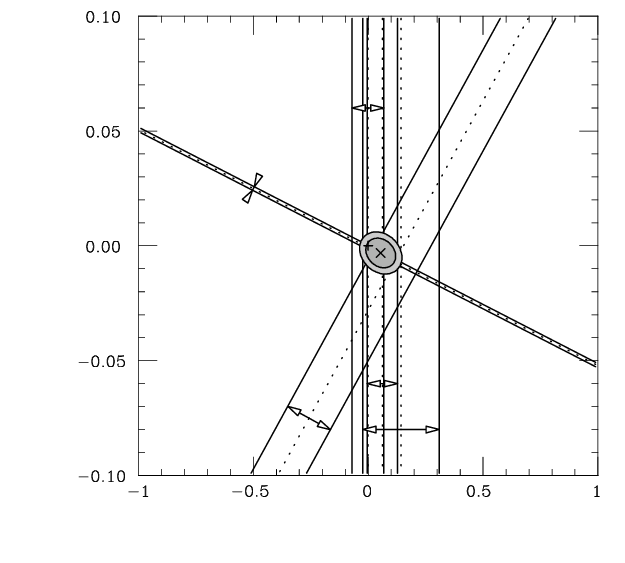}
\end{picture}
\caption{Constraints in the $\dhdR$--$\dhbR$ plane from various observables.}
\label{FIG35}
\end{figure}

%%%%%%%%%%%%%%%%%%%%%%%%%%%%%%%%%%%%

\begin{figure}[p]
\centering
\unitlength=1cm
\begin{picture}(12,10)
\unitlength=1mm
%\zahyou{12}{10}
\put(59,5){$\dhbR$}
\put(1,53){$\dhbLH$}
\put(86,63){$R_b$}
\put(47,27){$R_c$}
\put(43,70){$A_{\rm FB}(b)$}
\put(80,25){$A_b$}
\epsfbox[0 0 300 270]{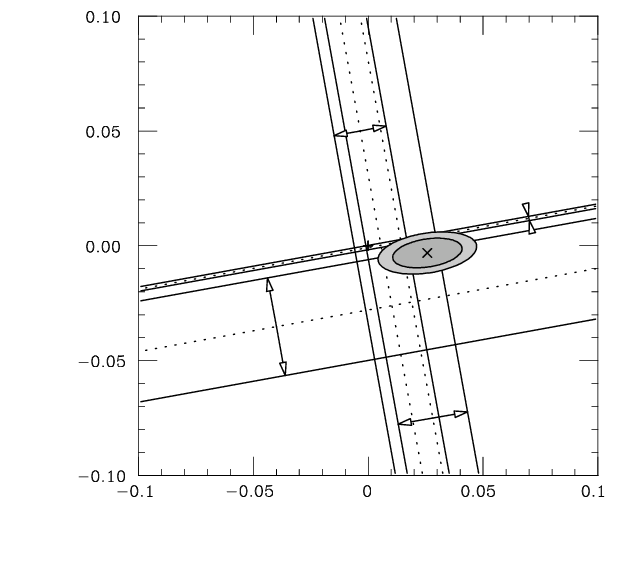}
\end{picture}
\caption{Constraints in the $\dhbR$--$\dhbLH$ plane from various observables.}
\label{FIG45}
\end{figure}

%%%%%%%%%%%%%%%%%%%%%%%%%%%%%%%%%%%%%%%%%%%%%%%%%%%%%%%%%%%%%%%%%%%%%%%%%%%%
\narrowtext

%%%%%%%%%%%%%%%%%%%%%%%%%%%%%%%%%%%%%%%%%%%%%%%%%%%%%%%%%%%%%%%%%%%%%%%%%%%%%
\end{document}